\documentclass[fleqn,usenatbib]{mnras}

\usepackage{newtxtext,newtxmath}
\usepackage[T1]{fontenc}
\usepackage{ae,aecompl}
\usepackage{graphicx}	
\usepackage{amsmath}	
\usepackage{amssymb}	

\usepackage{xcolor}
\usepackage{multirow} 

\newcommand{\msun}{{\rm M}_\odot}
\newcommand{\zsun}{{\rm Z}_\odot}

\newcommand{\code}[1]{\textsc{#1}}

\def\S*{$\Sigma_{\rm SFR}$}

\def\HI{\hbox{H~$\scriptstyle\rm I\ $}}

\def\Oj{O$^+$}
\def\Cj{C$^+$}

\def\cii{[CII]$_{158\mu}$}
\def\oia{[OI]$_{63\mu}$}
\def\oib{[OI]$_{146\mu}$}
\def\oiii{[OIII]$_{88\mu}$}
\definecolor{apcolor}{HTML}{b3003b}

\usepackage{etoolbox}
\makeatletter 
\patchcmd\@combinedblfloats{\box\@outputbox}{\unvbox\@outputbox}{}{\errmessage{\noexpand patch failed}}
\makeatother

\graphicspath{{Pictures/}}

\title[FIR line emission from simulated galaxies]{Predicting FIR lines from simulated galaxies}

\author[A. Lupi et al.]{Alessandro Lupi,$^{1}$\thanks{E-mail: alessandro.lupi@sns.it}
Andrea Pallottini,$^{1}$
Andrea Ferrara,$^{1}$
Stefano Bovino,$^{2}$\newauthor
Stefano Carniani$^{1}$,
and Livia Vallini$^{3}$
\\
$^{1}$Scuola Normale Superiore, Piazza dei Cavalieri 7, Pisa IT-56126, Italy\\
$^{2}$Departamento de Astronom\'ia, Facultad Ciencias F\'isicas y Matem\'aticas, Universidad de Concepci\'on,\\ Av. Esteban Iturra s/n Barrio Universitario, Casilla 160, Concepci\'on, Chile\\
$^{3}$Leiden Observatory, Leiden University, PO Box 9500, 2300 RA Leiden, The Netherlands \\
}

\date{Accepted XXX. Received YYY; in original form \today}
\pubyear{2020}

\begin{document}
\label{firstpage}
\pagerange{\pageref{firstpage}--\pageref{lastpage}}
\maketitle


\begin{abstract}
Far-infrared (FIR) emission lines are a powerful tool to investigate the properties of the interstellar medium, especially in high-redshift galaxies, where ALMA observations have provided unprecedented information. Interpreting such data with  state-of-the-art cosmological simulations post-processed with \code{cloudy}, has provided insights on the internal structure and gas dynamics of these systems. However, no detailed investigation of the consistency and uncertainties of this kind of analysis has been performed to date. Here, we compare different approaches to estimate FIR line emission from state-of-the-art cosmological simulations, either with \code{cloudy} or with on-the-fly non-equilibrium chemistry. We find that \cii~predictions are robust {to the model variations we explored}. [OI] emission lines{, that typically trace colder and denser gas relative to \cii,} are instead model-dependent, as these lines are strongly affected by the thermodynamic state of the gas and non-equilibrium photoionisation effects. For the same reasons, [OI] lines  represent an excellent {tool to constrain emission models, hence f}uture observations targeting these lines will be crucial. 
\end{abstract}



\section{Introduction}
In the last decades, deep high-resolution observations in the optical/near-infrared band have allowed us to start probing the high-redshift Universe and the Epoch of Reionisation, providing important insights on galaxy formation and evolution processes. 
However, to get a proper characterisation of the first galaxies, in particular the properties of the interstellar medium (ISM) at these redshifts, detailed spectral information is necessary. Low redshift observations have shown that fine structure lines in the far-infrared band -- like \cii, \oia, and \oiii~-- represent good tracers of the interstellar medium (ISM) thermodynamic state, and of the star formation process \citep{delooze14,herreracamus15}.
These lines are typically unaffected by dust, unlike the ultraviolet (UV) emission, and can be equally bright in the high-redshift Universe \citep[e.g.][]{carniani18,hashimoto19}. At high-redshift, FIR lines are shifted to the sub-mm band, where {they} are accessible by ALMA. In the last few years, ALMA observations have given us unprecedented data at extremely high resolution, providing important information about the ISM properties and kinematics of high-redshift sources.

Theoretically, several models have been developed to investigate the properties of high-redshift systems. While many of them focused on the relative impact of stellar feedback \citep[e.g.][]{pallottini17a,trebitsch18,ma18,katz17,katz19}, the chemical composition \citep{maio:2015,arata20}, and the role of active galactic nuclei \citep{trebitsch19,lupi19b}, other studies tried to more accurately describe how the FIR emission is affected by different gas conditions, \citep[e.g.][]{vallini13,olsen15}, also considering the impact of the cosmic microwave background and the metallicity on the powered emission \citep{vallini15,olsen17}, internal structure of molecular clouds (\citealt{vallini18}; Decataldo et al. in prep), the importance of the radiation field \citep{vallini:2017,pallottini19,arata:2019}, and possible bias in the comparison between theory and simulations \citep{kohandel:2019,lupi19b}.

In the majority of these studies, emission lines are estimated via post-processing with \code{cloudy} \citep{ferland98}, that provides {a very} accurate treatment of atomic and molecular microphysics, accounting for impinging radiation field and different ISM conditions. However, because of its complexity and computational cost, applying \code{cloudy} calculations on-the-fly in hydrodynamic simulations is not currently feasible, and cheaper approximate approaches have been developed, as machine learning techniques \citep{katz19} or multi-dimensional interpolation of tabulated emission lines \citep{pallottini19}. Moreover, \code{cloudy} assumes photo-ionisation equilibrium conditions, which does not always give a good representation of the real thermodynamic state of the ISM, especially for warm and cold gas \citep{bovino16}, and can produce less accurate heating/cooling rates for the gas in galaxies \citep{richings14,richings16}.

An alternative way to model emission lines is the direct inclusion of non-equilibrium calculations within simulations, providing all the necessary chemical abundances at every time. The abundances and thermodynamic state of the gas can then be used to compute the emission lines {by estimating the excitation level population of each ion assuming a statistical equilibrium among the levels}. This approach ensures a better representation of the evolution of the ISM, thanks to the consistent coupling of chemistry and thermodynamics. The time-dependent species abundances and cooling/heating processes thus ensure a strong interplay with all the sub-grid physics models included in the simulation. However, because of its complexity, this approach is currently feasible only for simplified chemical networks with a limited number of species \citep[e.g.][]{capelo18,lupi20}. 

{A crucial role in the emission line modelling is the temperature considered for each cell, that can be directly taken from the simulation, hence without considering any sub-resolution structure, as typically done in \code{krome} and in \citet{katz19}, or allowing for a depth-dependent distribution computed according to photo-ionisation equilibrium \citet{pallottini19}.}

In this study, we address how robust the FIR emission line prediction obtained in simulations is, depending on the photo-ionisation equilibrium and thermal state assumptions for the gas, by comparing post-processing \code{cloudy} calculations and on-the-fly non-equilibrium chemistry. To achieve this goal, we employ a state-of-the-art high-resolution zoom-in cosmological simulation of a typical star forming galaxy at $z=6$; the simulation includes a non-equilibrium chemical network, a physically-motivated star formation model, and stellar feedback by supernovae, winds, and radiation, the latter evolved through on-the-fly radiative transfer calculations.
The paper is organised as follows: in Section~\ref{sec:setup} we describe the numerical setup and the sub-grid models employed; in Section~\ref{sec:results} we present our simulation results, {and} in Section~\ref{sec:emission_lines} we discuss the emission line properties. {Then, in Section~\ref{sec:caveats} we discuss the caveats of the study and in Section~\ref{sec:conclusions} we finally} draw our conclusions.

\section{Numerical setup}\label{sec:setup}

We perform a cosmological zoom-in simulation targeting a $M_{\rm vir}\sim 3\times 10^{11}\,\msun$ halo at $z=6$ using the hydrodynamic code \code{gizmo} \citep{hopkins15}, descendant of \code{gadget3} and \code{gadget2} \citep{springel05}, employing the meshless-finite-mass method. Gravity is solved via a TreePM approach, with the maximum spatial resolution set by the Plummer-equivalent gravitational softening (kept fixed across the run) at 30~pc for dark matter (DM) and 10 pc for stars. For gas particles/cells, we employ adaptive softenings, with a minimum softening of 3~pc. The mass resolution is $\sim 10^5\,\msun$ for DM and $2\times 10^4\,\msun$ for gas and stars.
The initial conditions are created with \code{music} \citep{hahn13} assuming the \citet{planck16} cosmological parameters, where $\Omega_{\rm m} = 0.308$, $\Omega_{\rm b}=0.0481$, $\Omega_{\Lambda}=0.692$, $H_0=67.74\rm\, km\, s^{-1}\, Mpc^{-1}$, $\sigma_8 = 0.826$, and $n_{\rm s} =0.9629$.

The simulation presented here is performed employing the sub-grid model described in \citet{lupi20}, with the addition of the M1 on-the-fly radiation transport \citep{levermore84,hopkins20rad}.

\subsection{Sub-grid modelling}

Here, we recall the sub-grid models employed in our run, and describe the coupling between chemistry and radiation. 

\begin{itemize}
    \item \emph{Radiative cooling and chemistry}: we employ the non-equilibrium chemistry library \code{krome} \citep{grassi14}. Our network follows sixteen chemical species, i.e. H, H$^-$, H$^+$, H$_2$, H$_2^+$, He, He$^+$, He$^{++}$, C, C$^+$, O, O$^+$, Si, Si$^+$, and Si$^{++}$, properly taking into account the metal contribution {of these species} to gas cooling for $T<10^4$~K (instead of equilibrium tables), and the ultra-violet background (UVB) by \citet{haardt12}, as in \citet{capelo18} and \citet{lupi20}. {Our network includes photoheating (see Section~\ref{sec:photo}), H$_2$ UV pumping, Compton cooling, photoelectric heating, atomic cooling (from both primordial and metal species), H$_2$ cooling, and chemical heating and cooling. The contribution to metal cooling from other species/ionisation states not included in our network is only considered at $T>10^4$~K, where we employ equilibrium tables, whereas their effect is neglected at lower temperature, where in any case it is not expected to be important \citep[see, e.g.][]{glover07,richings14,bovino16}}. Photo-chemistry is implemented accounting for the photon flux within each cell in ten different radiative bins spanning the range 0.75-1000~eV \citep{lupi18}.
    \item \emph{Star formation (SF)}: SF is modelled via a stochastic approach with the efficiency {
    \begin{equation}
        \varepsilon_{\rm SF}=\varepsilon_{\rm SF,0} \exp\left(-1.6\frac{t_{\rm ff}}{t_{\rm dyn}}\right),
    \end{equation}
    } computed on a cell/particle basis as a function of {the local free-fall time $t_{\rm ff}=\sqrt{3\pi/(32{\rm G}\rho_{\rm gas})}$ and dynamical time $t_{\rm dyn} = h/(2\sigma_{\rm eff})$, where G is the gravitational constant, $\rho_{\rm gas}$ is the cell/particle gas density, $h$ is the effective size of the cell and $\sigma_{\rm eff} = \sqrt{c_s^2+\sigma_{\rm v}^2}$ the total thermal+turbulent support, identified by the sound speed $c_s$ and the velocity dispersion $\sigma_{\rm v}$, respectively.} This prescription, already employed in \citet{lupi20} and based on the theoretical model by \citet{padoan12}, assumes that every particle/cell in the simulation represents a entire molecular cloud (or a large portion of it) described by a Log-Normal probability distribution function, {whose parameters (Mach number, mean density, and virial parameter) are determined on-the-fly according to the local gas properties \citep{lupi19b,lupi20}. As in \citet{semenov17}, the local efficiency within the cloud is assumed to be {$\varepsilon_{\rm SF,0}=0.9$} at the pre-stellar core scale.}
    \item \emph{Stellar winds and supernova feedback}: due to the limited mass resolution, every stellar particle is assumed to represent an entire population following a \citet{kroupa01} Initial Mass Function (IMF). Energy, mass, and metals are released by SNe in discrete events \citep{hopkins18b,lupi19a} of $10^{51}$~erg each, where thermal energy and momentum are directly added to the gas according to the results by \citet{martizzi15}. As in our previous studies \citep{lupi18,lupi19a,lupi19b}, metal injection by SNe is computed by scaling O and Fe abundances to a total $Z$ according to the solar ratios \citep{asplund09}, that give $M_{Z} = 2.09 M_{\rm O} + 1.06 M_{\rm Fe}$ \citep{kim14}. For stellar winds, we do not consider any additional metal production relative to the stellar progenitor metal fraction, and dump feedback as thermal energy only, in addition to the initial ejecta momentum.\footnote{In order to properly conserve species abundances for the chemical evolution, metal species (and ions) abundances are re-scaled, for active gas particles, to match the updated total metallicity of the particle.}
     \item \emph{Stellar radiation}: We model stellar radiation assuming the up-to-date \citet{bruzual03} stellar population synthesis models for a Kroupa IMF. The spectra are sampled in ten radiation bins as a function of stellar age and metallicity, with each stellar particle representing an entire stellar population.
 \end{itemize}

\subsection{Radiation transport and photo-chemistry}
\label{sec:photo}
{In our simulation, we employ a moment-based radiation transport (RT) scheme based on the widely used M1 closure scheme {\citep{levermore84,aubert:2008,skinner13,rosdahl13,hopkins20rad}}, where the first and second momenta of the RT equation are solved assuming a purely local form of the Eddington tensor \citep{levermore84}. This scheme results in hyperbolic equations, that can be solved with the Godunov-like method also employed for hydrodynamics \citep{aubert:2008}. To make the run feasible, we employ the {\it reduced speed of light approximation} \citep{gnedin01} with a reduction factor $f_c=0.001$.\footnote{{We note that our choice for $f_c$ corresponds to a speed of light comparable to the gas velocities in high redshift galaxies. Although $f_c=0.01$ would have represented a more accurate choice, this would have implied a ten times higher computational cost of the run, which was already high because of the non-equilibrium chemistry coupling. Moreover, as discussed in \citep{hopkins20rad}, the effect of $f_c$ was moderate in the galaxy ISM, with values lower than $\tilde{c}=1000\rm\, km\, s^{-1}$ simply resulting in a slightly more effective radiative feedback.}}}

In its standard implementation within \code{gizmo}, the signal speed eigenvalues are derived under the HLL approximation, and are obtained from a fit to the two-dimensional table created by \citet{rosdahl13}. Here, we opt instead for the more accurate (and purely analytical) derivation reported in \citet{skinner13}, that gives
\begin{equation}
    \frac{\lambda_{1,3}}{\tilde{c}} = \left\{\mu f \pm \left[ \frac{2}{3}\left(\xi^2 -\xi\right) + 2\mu^2\left(2-f^2 - \xi\right)\right]^{1/2}\right\}/\xi,
\end{equation}
where $\tilde{c}=f_{c}c$, $c$ is the speed of light, 
$f=|F_\nu|/(\tilde{c}E_\nu)$ is the reduced flux, $\mu=\cos{\theta}$ is the angle between the radiation flux and the face vectors, and $\xi=\sqrt{4-3f^2}$.

\begin{figure}
    \centering
    \includegraphics[width=\columnwidth]{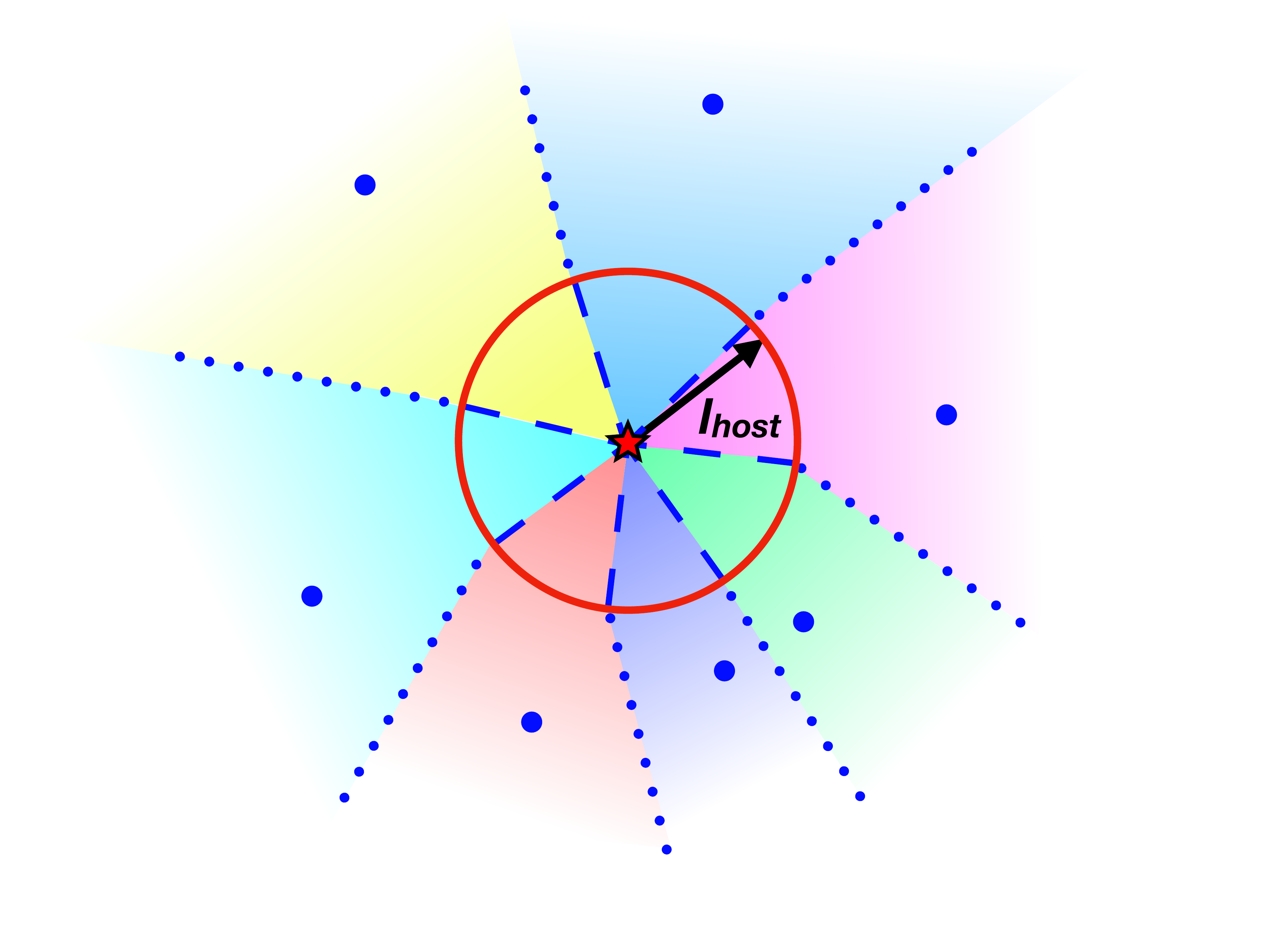}
    \caption{Schematic view of the sub-grid absorption applied during the photon injection step, to account for unresolved absorption in the simulation and prevent the artificial excess of escaping radiation. The red circle represents the `virtual host cell' around the source, with size $l_{\rm host}$; {the different colours correspond to the region associated to each neighbouring gas particle $j$, that determines the fraction of the total luminosity dumped on that particle itself}. 
    \label{fig:scheme}
    }
\end{figure}

At every time-step, all active stellar particles inject photons into the nearest $N_{\rm ngbs}\approx 64$ gas neighbours\footnote{The number of neighbours for stars is twice that of the gas, and this choice is made to ensure that the region around the source is well sampled \citep[see, e.g.,][for details]{lupi19a}}, weighted by a parabolic kernel as a function of distance, as it is done in the public version of \code{gizmo}.
Because of the intrinsic particle-based nature of \code{gizmo}, radiation cannot be injected in a single `host cell' as commonly done in grid-based codes, but is spread over a discrete number of neighbours. This process could in principle artificially boost the escape of radiation from high-density regions, and overestimate the radiative feedback of the star, especially if the farthest neighbours are optically thin to radiation. In order to prevent this issue, and account for unresolved absorption, for every energy bin, we attenuate the photons injected accounting for the column density between the source and the target cell $j$ as
\begin{equation}
\Delta N'_{\gamma,j} = \Delta N_{\gamma,j} \exp{(-\kappa_j\rho_j r_{{\rm eff},j})},
\end{equation}
where $\kappa_j$ and $\rho_j$ are the neighbour cell opacity and density and $r_{{\rm eff},j} = \max\{l_{\rm host},r_{\star,j}\}$, with $l_{\rm host}$ is the size of the `virtual host cell' around the stellar source, and $r_{\star,j}$ is the separation between the target cell and the source. As in \citet{hopkins18b} and \citet{lupi19a}, we define $l_{\rm host}=[3H_{\rm host}/(4\pi N_{\rm Ngb})]^{1/3}$, where $H_{\rm host}$ is the kernel size of the star encompassing $N_{\rm Ngb}=64$ neighbours. A schematic view of this approach is reported in Fig.~\ref{fig:scheme}.
In addition to this correction, we also account for the unresolved radiation pressure on the neighbours, by imparting a kick defined as\footnote{An unresolved radiation pressure term is already implemented in the public version of \code{gizmo}, but uses a different definition for the absorption length, i.e. $r_{{\rm eff},j}=2\max\{2h_j,H_{\rm host}\}$, with $h_j$ the effective size of the $j$-th cell.}
\begin{equation}
    m_j\Delta \mathbf{v} = \frac{\Delta E_{\gamma,j}}{c} [1-\exp{(-k_j\rho_jr_{{\rm eff},j})}],
\end{equation}
where $\Delta E_{\gamma,j} = \Delta N_{\gamma,j} \langle E_{\gamma}\rangle$ is the photon energy injected and $\langle E_{\gamma}\rangle$ the average photon energy per bin.

To couple radiation with chemistry, we follow the same approach described in \citet{lupi18}, where {photoionisation rates} and photo-heating are computed in \code{krome} from the photon fluxes within each cell of the simulation, defined for every radiation bin $j$ as $F_{\gamma,j} = \tilde{c}E_{\gamma,j}/V_{\rm cell}$, with $E_{\gamma,j}$ the 
$j$-th bin photon energy, and $V_{\rm cell}=h_{\rm cell}^3$ the cell volume. 
While the opacity resulting from the abundance of the species included in the chemical network is consistently treated by \code{krome}, and used to properly attenuate radiation in the RT scheme, dust shielding and H$_2$ self-shielding are included only in the chemistry calculations assuming an effective absorption scale equivalent to the Jeans length capped at 40~K \citep{safranekshrader17} for each gas cell/particle \citep{lupi19b}. 

{We further assume the \textit{on-the-spot approximation}, and neglect any photons produced by the gas during the cosmic evolution.}

 \section{Results}\label{sec:results}
We now present our results, and discuss the uncertainties in the predicted emission lines for our target galaxy.
For all the analyses reported here, we have identified the $M_{\rm vir}\sim 3\times 10^{11}\,\msun$ target halo using \code{amiga halo finder} \citep{knollmann09}, and we have considered all the particles/cells within 20 per cent of the halo virial radius, in order to exclude any contamination by satellites. As an example, at $z\sim 6$, the halo virial radius is about $28.4$~kpc, hence we consider particles within $5.7$~kpc only.

\subsection{Galaxy evolution across cosmic time}
\begin{figure*}
    \centering
    \includegraphics[width=\columnwidth]{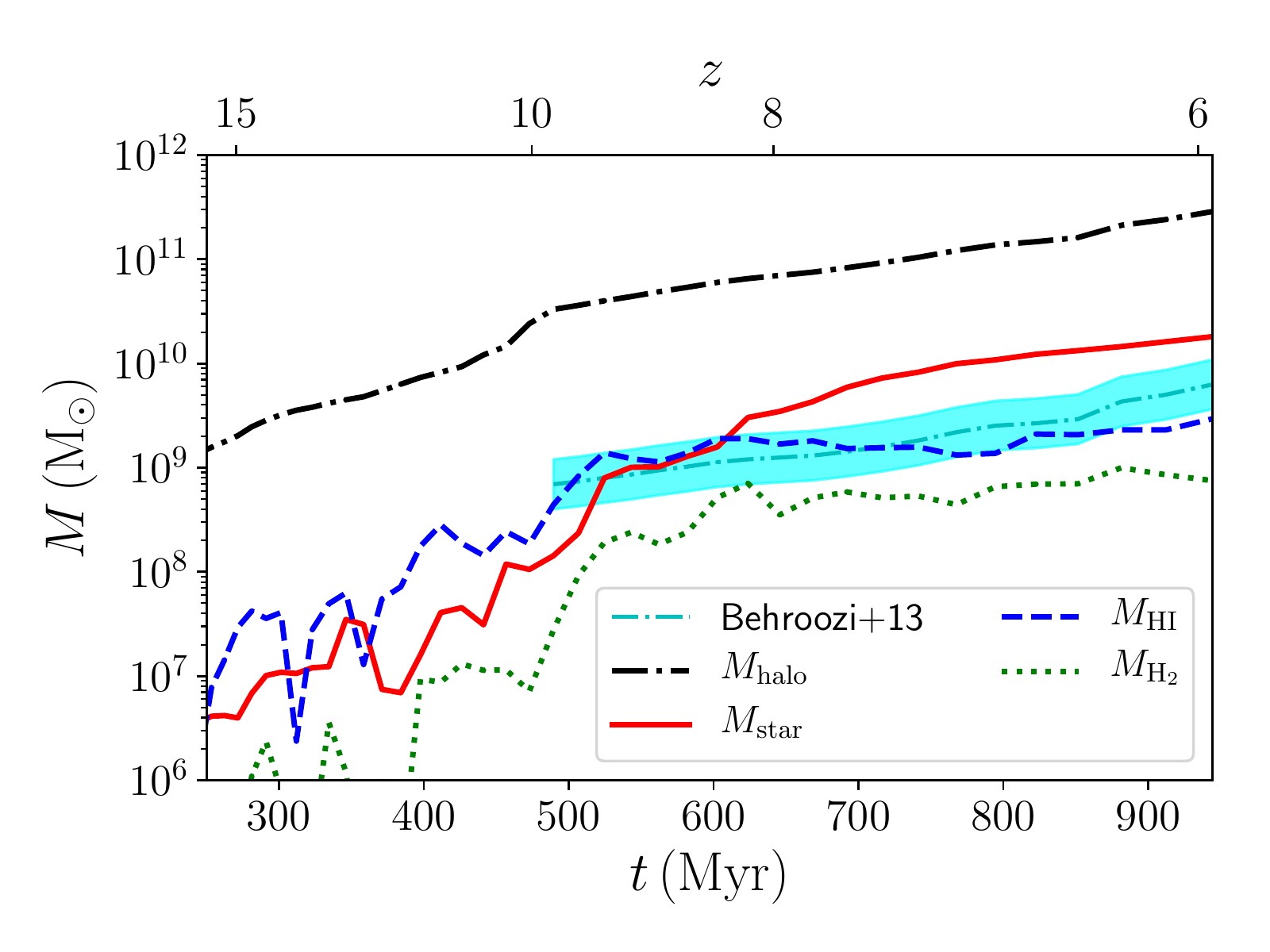}
    \includegraphics[width=\columnwidth]{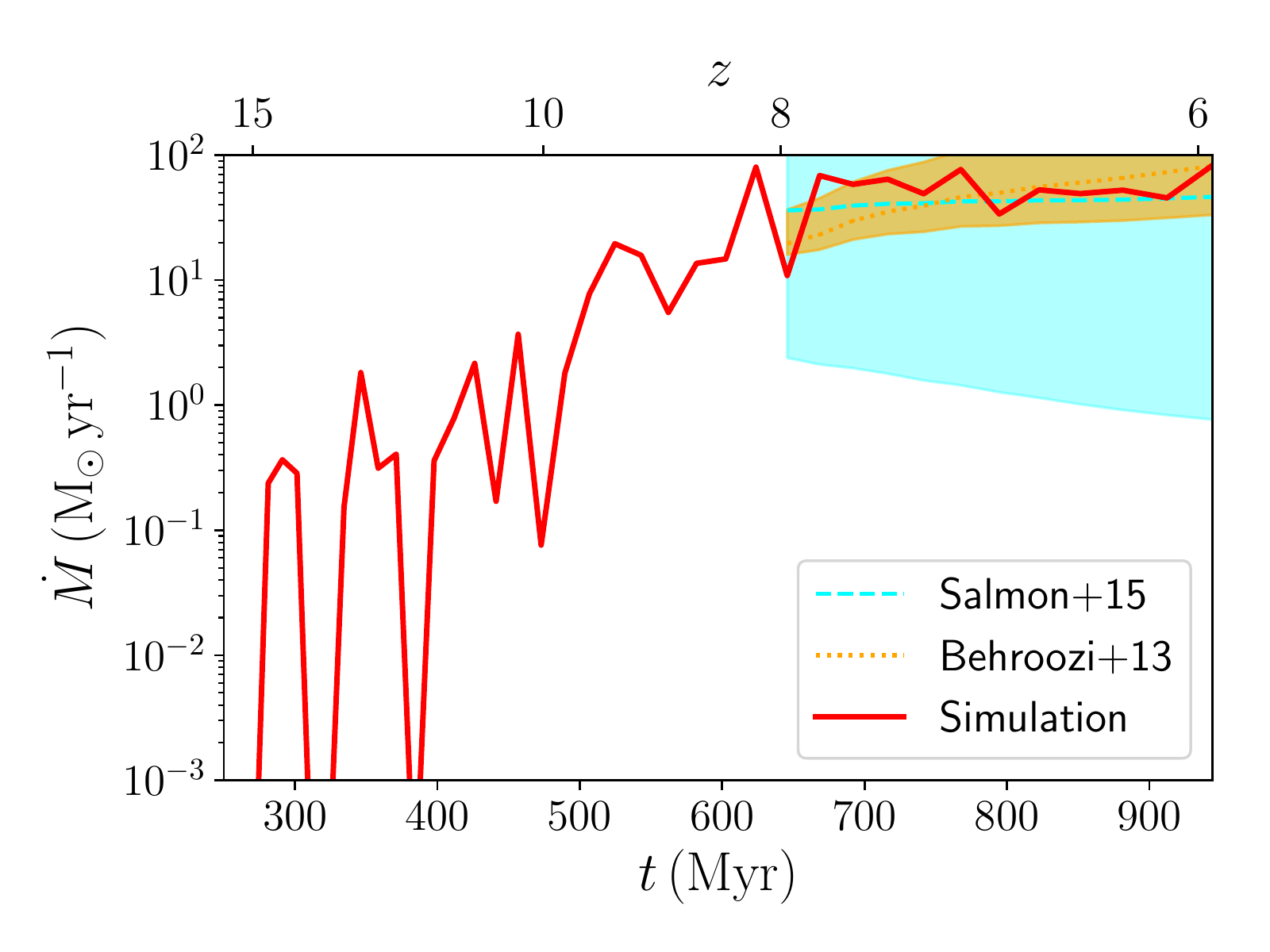}
    \includegraphics[width=\columnwidth]{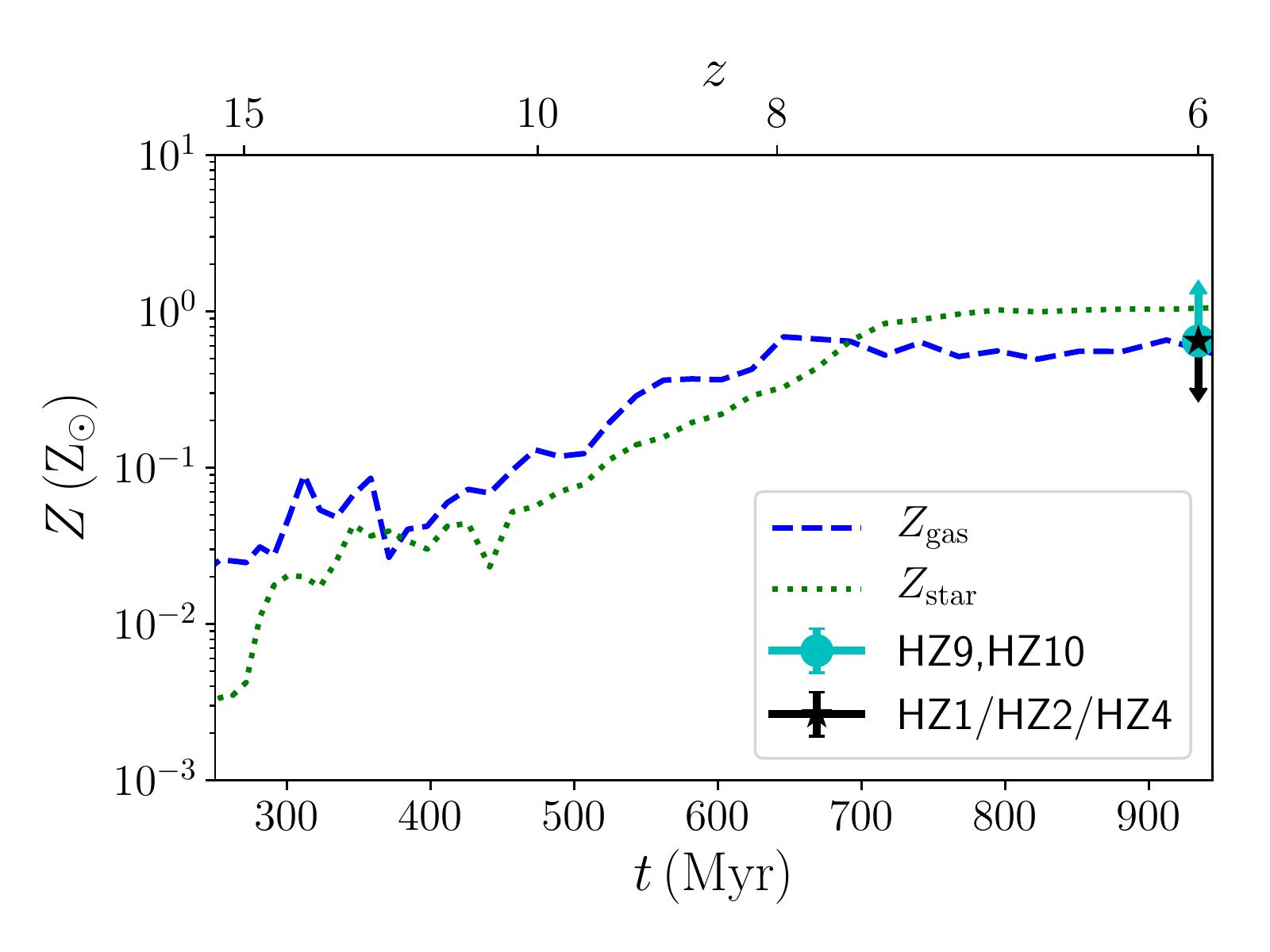}
    \includegraphics[width=\columnwidth]{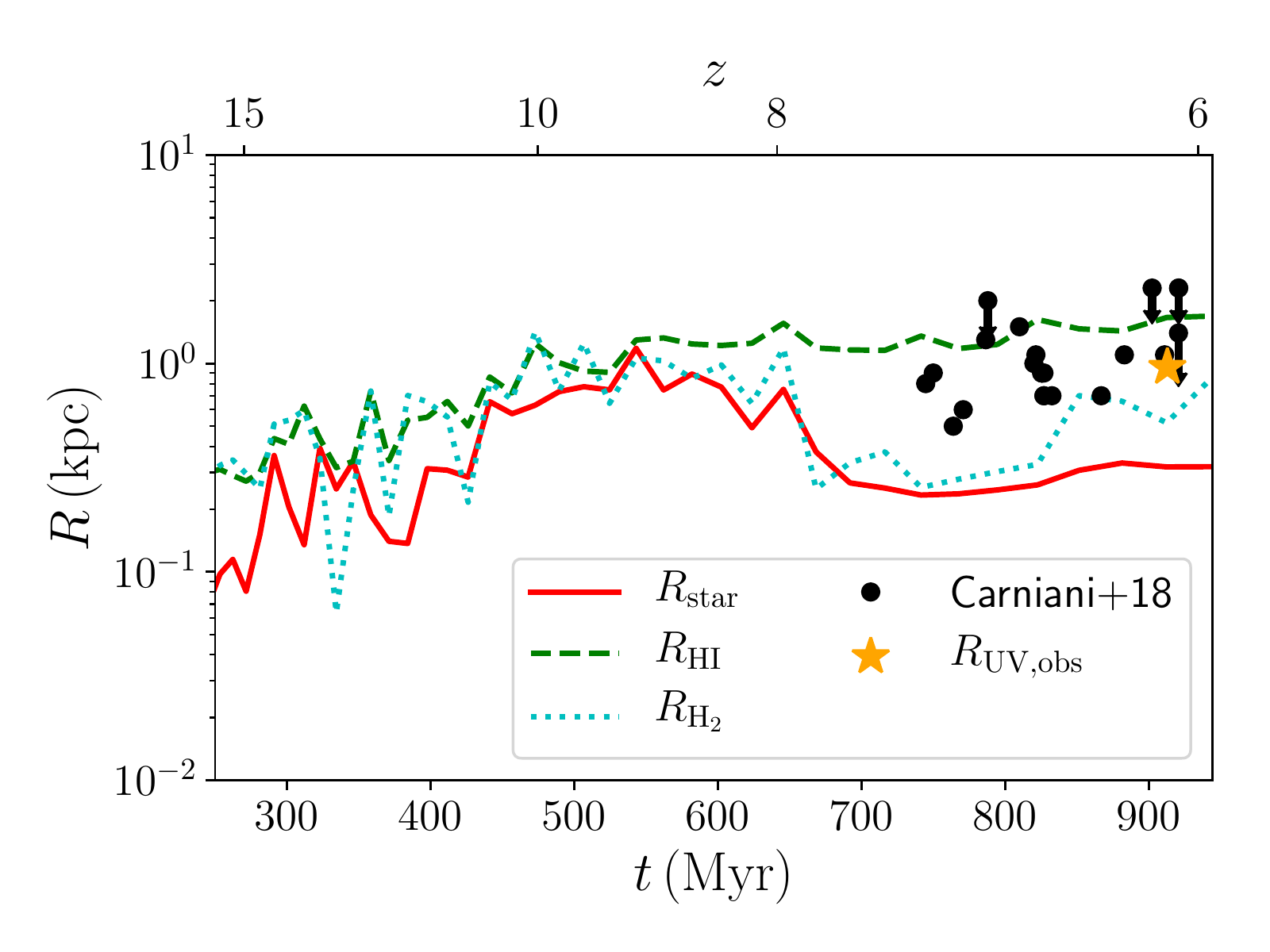}
    \caption{Redshift evolution of galaxy properties. {\bf Top-left panel}: halo mass (black dot-dashed line), stellar mass (red solid), neutral hydrogen mass (blue dashed line), and molecular hydrogen mass (green dotted). {The cyan shaded area corresponds to the model by \citet{behroozi13}.} 
    {\bf Top-right}: Star formation rate (SFR, red solid line){, compared with the empirical model by \citet[][orange shaded area]{behroozi13} and the observational constraints by \citet[][cyan shaded area]{salmon15}}. {\bf Bottom-left}: gas (blue dashed) and stellar (green dotted) metallicity{, compared with the high-redshift observations by \citet{faisst17}}. {\bf Bottom-right}: Half-mass radii for stars (red solid line), HI (green dashed), and H$_2$ (cyan dotted). {The black dots correspond to the data by \citet{carniani18}, and the orange star to a crude estimate of the effective size of our galaxy when dust absorption is considered.}
    }
    \label{fig:mass}
\end{figure*}

\begin{figure*}
    \centering
    \includegraphics[width=\textwidth,trim=2cm 1cm 2cm 1cm,clip]{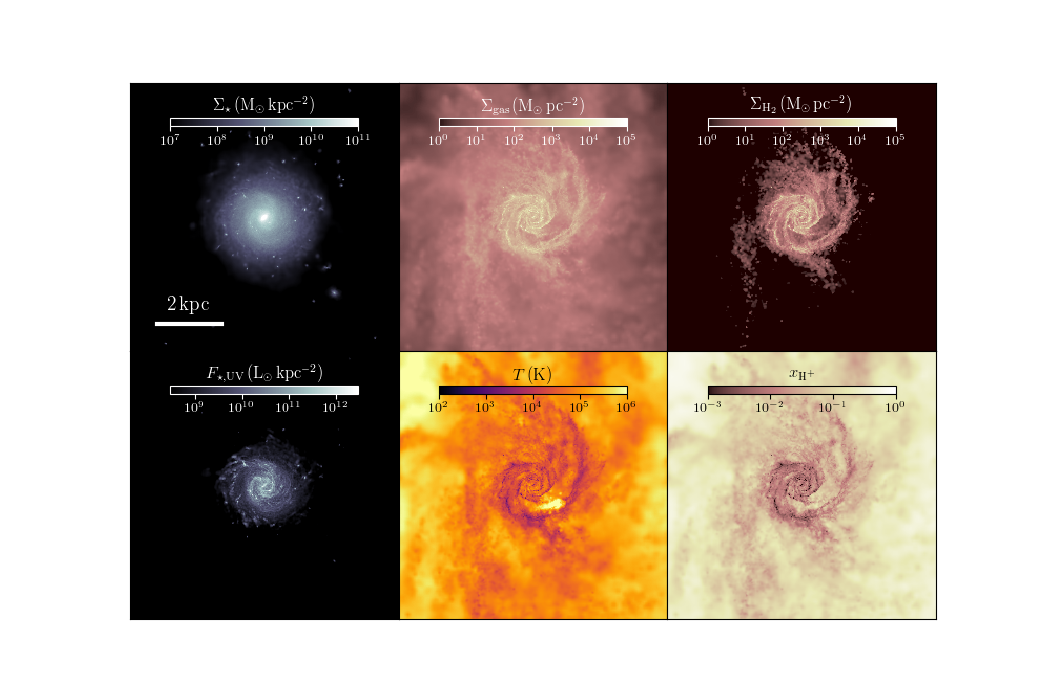}
    \caption{Maps of the key properties of our simulated galaxy at $z\sim6$. The top-left panel shows the stellar surface density, whereas the intrinsic FUV flux from the stellar spectra, tracing young stars, is shown in the bottom-left panel. In the second column, the total gas column density is reported in the top one, and {the line-of-sight averaged temperature of the galaxy in the bottom one. Finally, in the top-right panel, we report the H$_2$ column density,} and in the bottom-right one the ionised hydrogen fraction $x_{\rm H^+}$. By comparing the FUV emission with H$_2$, we notice that H$_2$ traces well the distribution of young stars forming within molecular clouds, whereas the global stellar distribution in the galaxy (leftmost panel), still dominated by older stars, is more concentrated. Compared to the molecular component, the neutral and ionised components are more diffuse, extending up to several kpc, as highlighted by the temperature and the $x_{\rm H^+}$ maps.}
    \label{fig:maps6}
\end{figure*}

In the top-left panel of Fig.~\ref{fig:mass}, we show the redshift evolution of the mass for different components. 
The stellar mass build-up starts about 200~Myr after the Big Bang, but at a moderately slow pace until $z\sim 12$, when the halo exceeds $M_{\rm halo}\sim 10^{10}\,\msun$ and SNe stop evacuating most of the gas. This is reflected in the initially bursty SFR shown in the top-right panel, that becomes steadier and converges to about $40-50\,\msun\rm\, yr^{-1}$ below $z\sim 10-9$. At very high redshift, gas makes up for most of the baryons, with neutral hydrogen representing the most abundant element, and molecular hydrogen making up for less than 5$\%$. Below $z=9$, stars start to dominate, and the gas-to-star ratio settles around 30~per cent (corresponding to a gas fraction $f_g=0.23$), with neutral hydrogen contributing for about 40-50~per cent, and the molecular component up to 15~per cent \citep[see, also,][]{popping15}. While at $z\gtrsim9$ our galaxy is consistent with the empirical stellar-to-halo mass relation by \citet{behroozi13} and \citet{behroozi19}{, \footnote{We note that, above $z=8$, the relation is extrapolated.} reported as a cyan dashed line (best-fit) and shaded area (1-$\sigma$ uncertainty)}, at $z\sim 6$ our stellar mass ($M_{\rm star}=1.6\times 10^{10}\rm\, M_\odot$) is about 2--2.5 times above the relation, because of the weaker effect of SNe at larger halo masses, although still compatible within $3\sigma$.

The SFR, instead, is roughly constant around $50\rm\, M_\odot\, yr^{-1}$ below $z\sim 8$, in perfect agreement with the observational constraints by \citet{salmon15}, { shown as the cyan shaded area and dashed line,} for which $\dot{M}_{\rm star} \approx 45\rm\, M_\odot\, yr^{-1}$, and the empirical estimates by \citet{behroozi13},{ shown as the orange shaded area and dotted line,} for which $\dot{M}_{\rm star} \approx 72\rm\, M_\odot\, yr^{-1}$. 

In the bottom-left panel, we also show the stellar and gas metallicity of the galaxy. While the stellar metallicity increases almost monotonically with time, reaching solar values around $z=7.5$, gas exhibits stronger fluctuations, due to the competing effect of SN enrichment and pristine gas inflows. Around $z=7-6$, the average gas metallicity has saturated around $Z_{\rm gas}=0.5\,\rm Z_\odot$, whereas stars are twice more metal-rich relative to gas. The gas metallicity we get is consistent with the estimates {(upper/lower limits)} for high-redshift galaxies observations \citep[e.g.][]{faisst17,carniani18}{, reported as cyan (black) lower (upper) limits,\footnote{The data have been shifted to $z=6$ from $z\sim 5.6$ only for illustrative purposes.}} and similar simulations \citep[e.g.][]{pallottini17b}.

Finally, in the bottom-right panel, we report the size evolution of the different galaxy components. The half-mass radius for the stellar component is shown as a red solid line, \HI as green dashed one, and H$_2$ as a cyan dotted one. At early times, only a few clumps exist, where H$_2$ is embedded within neutral gas. This results in a comparable size for the two components, whereas stars only form in the densest region at the centre of the system, resulting in a more concentrated distribution. At $z\lesssim 11$, the occurrence of a major merger results in an apparent rapid size increase, with the peak at $\sim 800$~pc at $z\sim 10$, followed by a sharp drop in size when the merger ends and the galaxy remnant settles ($z\sim 8-9$). Below $z=8$, the galaxy exhibits a well-defined gaseous disc, mostly H$_2$ rich, where stars continuously form, whereas the remaining neutral gas is more extended, up to a few kpc. 
Around $z=6$, the stellar distribution remains quite compact, with a typical half-mass (and {intrinsic} half-light) radius of about 300-400~pc, a value only moderately smaller than the typically observed one for Lyman-break galaxies, { as shown by the black dots and upper limits \citep{carniani18} and, e.g., Figure 32 of  \citet{dayal18}. {We notice, however, that dust attenuation could affect the observed light profile of the galaxy, changing its apparent size. To show this effect, we report as an orange star the UV `observed' size resulting from a crude approximation of the dust attenuation to the FUV image of the galaxy at $z=6$ (bottom-left panel of Fig.~\ref{fig:maps6}), obtained via the following procedure: \textit{(i)} we first estimated the dust optical depth along the line-of-sight $\tau_{\rm d}= 0.5k_{\rm UV}\Sigma{\rm gas}f_{\rm d}$, where we assumed $k_{\rm UV}=4.2\times 10^4\rm\, cm^2\, g^{-1}$ as the dust opacity at 1500~\AA, $\Sigma_{\rm gas}$ is the gas column density, $f_{\rm d}={\rm D_\odot}(Z/{\rm Z_\odot})$ is the dust-to-gas ratio \citep[assuming D$_\odot$=0.00934 and Z$_\odot$=0.013;][]{grassi17}, and the factor 0.5 accounts for the half of the disc only; \textit{(ii)} we then attenuated the FUV flux with $\exp(-\tau_{\rm dust})$, and \textit{(iii)} finally fitted the resulting image via a 2D Gaussian kernel, extracting the half-light radius $R_{\rm UV,obs}$.}}

\subsection{Galaxy properties at $z=6$}

\begin{figure}
    \centering
    \includegraphics[width=0.9\columnwidth]{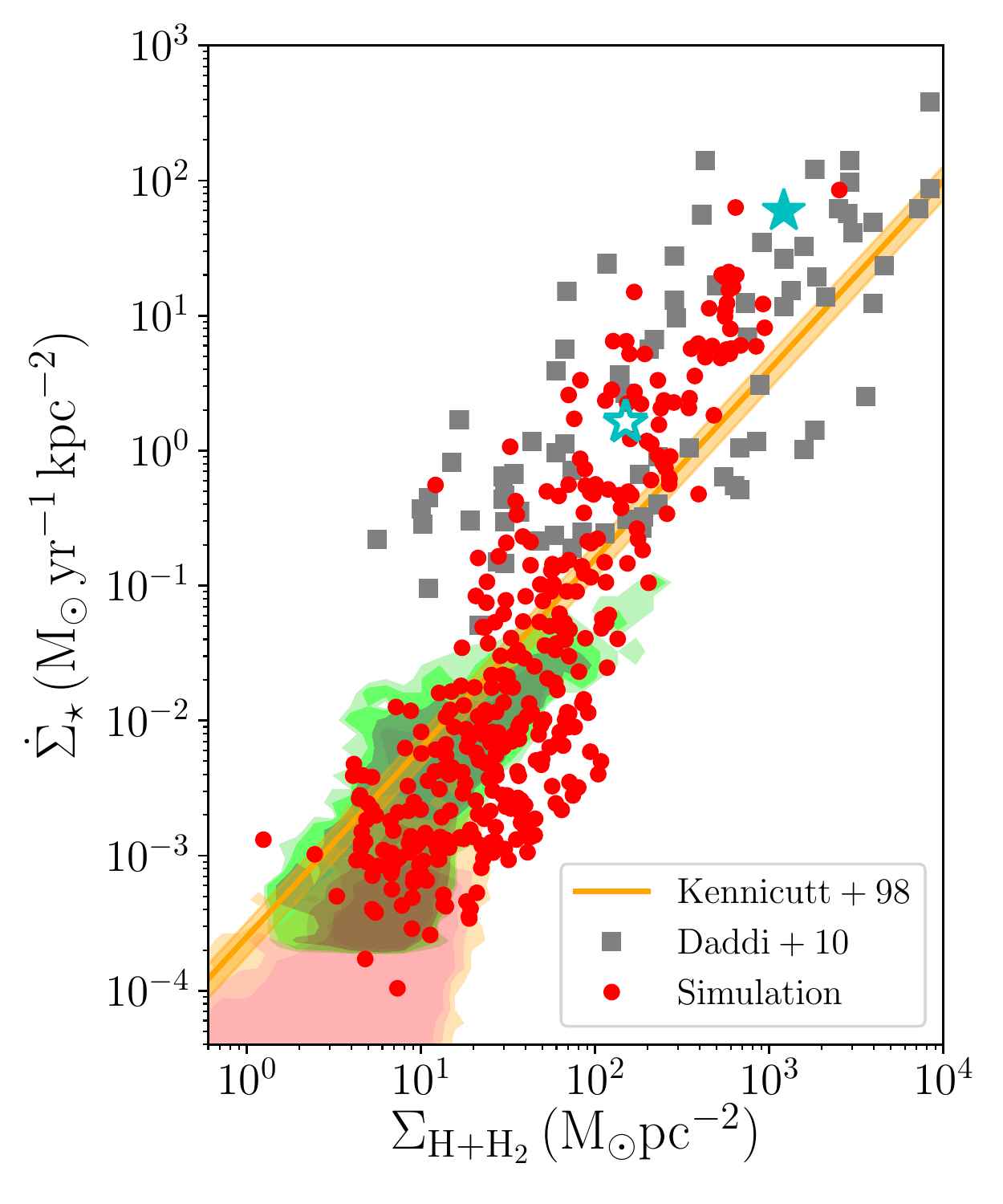}
    \caption{KS relation in the galaxy, shown as red points, compared with local data by \citet{bigiel10} (red/green contours), the best-fit by \citet{kennicutt98} (orange shaded area and solid line), and the SB/SMG/ULIRG data reported in \citet{daddi10}. The star symbols correspond to the average values for our galaxy, averaged over 2~kpc (empty cyan star) or within the H$_2$ half-mass radius for the gas and the stellar half-mass radius for stars (filled cyan star). The agreement with observations is very good across the entire range, with also the kink around $\Sigma_{\rm H+H_2}=10\,\msun\, {\rm pc}^2$ well reproduced. At high densities ($\Sigma_{\rm H+H_2}\gtrsim 30\,\msun\, {\rm pc}^2$), we observe the transition towards the SB region for a few patches, in perfect agreement with the data by \citet{daddi10}.}
    \label{fig:ks}
\end{figure}

Next, we focus on the properties of the galaxy at $z\sim 6$, with particular emphasis on the ISM.
In Fig.~\ref{fig:maps6}, we show the spatial distribution of the different galaxy components. The stellar distribution (top-left panel) exhibits a smooth disc structure, with well defined spiral arms. The galaxy is compact, with most of the stellar mass within 1~kpc (the total half-mass radius is about 300~pc, see bottom-right panel of Fig.~\ref{fig:mass}).
Using the intrinsic FUV emission, computed from the \citet{bruzual03} stellar spectra in the band $1350-1750$~\AA, we can identify star forming regions (bottom-left panel): young stars forming from the gaseous disc also settle on a disc, although the distribution is more clumpy, reflecting the clustering of stars forming within molecular clouds. With time, stellar radiation, winds, and SNe evacuate gas from the clouds, and the stellar clusters slowly disperse within the disc. 

Compared to local galaxies, high-redshift systems continuously accrete gas from the environment, and this provides new fuel to sustain SF (which in these systems can easily exceed 10~$\msun\,\rm yr^{-1}$). As shown by the total gas column density (first row, central panel), a significant amount of gas extends up to several kpc; such gas has either been expelled by stellar feedback or is flowing in from large-scale filaments. On the contrary, {the} main galaxy disc does not exceed 1--2~kpc. The H$_2$ distribution (central panel, second row) closely follows that of young stars, consistently with expectations that SF mostly occurs in cold and dense molecular clouds where H$_2$ is abundant. We stress that the H$_2$ distribution in our run is a natural byproduct of our simulation, despite we do not assume any dependence of SF on the H$_2$ abundance \citep{lupi18}. Finally, in the right panels, we report the average gas temperature along the line-of-sight (top panel), and H ionization fraction, $x_{\rm H^+}$ (bottom panel). Most of the gas in the spiral arms is cold, as expected for H$_2$-dominated gas, whereas the inter-spiral regions with lower density gas are warm/hot, as they are heated and ionized by stellar radiation. Outside the galactic disc, the gas is typically hotter, with warm filaments embedded within low-density regions with {high} temperatures, as also highlighted by the ionisation fraction that approaches unity.

One of the most fundamental correlations between galaxy properties is the Schmidt--Kennicutt (KS) relation \citep{schmidt59,kennicutt98}, that links the neutral and molecular gas reservoir to the effective SFR. In Fig.~\ref{fig:ks}, we show the KS relation in total gas from observations of different systems (normal local spirals as well as high-redshift or starburst galaxies) with our run. We show the best-fit by \citet{kennicutt98} (orange shaded area and solid line), the spatially-resolved local measures by \citet{bigiel10} (filled red/green contours), and the measures of starburst (SB), sub-millimeter (SMG), and ultra-luminous infrared (ULIRGs) galaxies reported in \citet{daddi10} (gray squares). {In our run (where the red points correspond to 0.3~kpc patches),} the low-density region matches the local data very well, also exhibiting the kink around $\Sigma_{\rm H+H_2}=10\,\msun\,\rm yr^{-1}$, whereas the higher density patches settle in the starburst regime \citet{daddi10}.
For completeness, we also show the average values for our galaxy as star symbols. The empty one corresponds to the average obtained at 2~kpc resolution, whereas the filled one is obtained as the ratio between the total H+H$_2$ mass (SFR) within 4~kpc and twice the surface of the region encompassing half of the H$_2$ (stars) mass, similarly to \citet{miettinen17}.\footnote{We note that, if the considered radii are significantly different, this choice can artificially bias the correlation, hence the use of similar radii to average both quantities, as usually done when the galaxy is resolved, is preferred.} In both cases, our galaxy matches the starburst data by \citet{daddi10}, further confirming the results obtained with the patches.
This suggests that, although high-$z$ galaxies are more likely to exhibit star-bursting phases, with SFRs well above the local relation, the spatially-resolved relation is not dissimilar from the local one, with lower density regions {($\Sigma_{\rm H+H_2}\lesssim 40\rm\, M_\odot\, pc^{-2}$)} perfectly consistent with nearby disc galaxies, and the densest regions within the disc {($\Sigma_{\rm H+H_2}\gtrsim 40\rm\, M_\odot\, pc^{-2}$)} reproducing low-redshift starbursts. Since most of the emission associated with newly formed stars comes from the densest regions, these early galaxies are likely to appear as starbursts, as found by \citet{vallini:2020}. However, because of the limited resolution and sensitivity of current observations, such an analysis cannot be performed on many sources, and a clear consensus about their location relative to the KS relation is still missing \citep[see, e.g.][]{pavesi19}.

\section{Testing FIR line predictions}\label{sec:emission_lines}

Now, we focus on the FIR emission from the galaxy at $z=6$, and investigate how the assumptions on the thermodynamic and ionisation state of the gas affect the estimated flux. This is crucial to assess the robustness of predictions, and how observations could help constraining the sub-grid prescriptions adopted in simulations \citep{olsen18}.

\subsection{Modelling emission lines}

\begin{table*}
    \begin{center}
    \caption{Differences between \code{cloudy} and \code{krome} in the method emission lines are estimated {in this work} and in the assumptions made for the chemical/thermodynamic state of the gas. {\it All the features reported here refer to single cell calculation at a specific time, and do not reflect how thermodynamics is evolved during the simulation}. In particular, CloudyFIX and CloudyVAR differ in how the temperature is evolved within the slab, either constant and based on the simulation outputs, in this way taking into account any dynamical effect (CloudyFIX), or let free to vary according to the radiation attenuation computed by \code{cloudy} (CloudyVAR).
    \label{tab:models}
    }
    \begin{tabular}{llll}
    \hline\hline
    Feature                 & Krome                                 & CloudyVAR                       & CloudyFIX$^\dag$ \\
    \hline
    Spatial structure       & Average properties (0D)               & slab (1D)                       & ~           \\
    Minimum size considered & as simulation resolution              & optically thin slices           & ~           \\
    Coupling with the
    simulation feedback     & Self-consistent                       & No                              & Partial     \\
    Chemical network        & 5 atoms and 1 molecule                & 30 atoms and multiple molecules & ~           \\
    Atomic levels hierarchy & Dominant levels only                  & Full                            & ~           \\
    Chemical state          & Full non-equilibrium                  & Ionisation equilibrium          &             \\
    Thermodynamic state     & Constant T                            & Variable T                      & Constant T  \\
    Effect of shocks        & Yes                                   & No                              & Yes         \\
    \hline\hline
    \end{tabular}
    \end{center}
\begin{flushleft}
    {\bf Notes}: $^\dag$Features not shown in the CloudyFIX column are the same as CloudyVAR.
\end{flushleft}
\end{table*}

FIR emission lines can be inferred from numerical simulations following different approaches. In particular, in this work we consider three different models, two based on \code{cloudy} \citep{ferland17}, i.e. model `CloudyFIX' and model `CloudyVAR', and one directly based on the chemical state of the gas in our simulation determined by \code{krome} (model `Krome'), as detailed in the following Sections. 

We stress that the Krome model relies on the self-consistent non-equilibrium chemical evolution within the cosmological simulation, where thermodynamics and chemistry are fully coupled, while this is not the case for CloudyFIX and CloudyVAR, that are simply applied to the outputs.

\subsubsection{Extracting FIR lines from KROME}\label{sub:sec:emission_krome}

Thanks to our non-equilibrium chemical network, we can directly follow the abundances of different ions in our simulation, that are self-consistently evolved during the run in a fully-coupled fashion together with the thermodynamic state of the gas. The abundances of each species are tracked for all cells in the simulation, and correspond to the average abundances within the cell.\footnote{{This implies that chemistry is solved assuming an optically thin slab, and photon absorption is then applied on the entire cell size to update the radiation flux transmitted to adjacent cells.}}  In particular, heating and cooling processes of the gas are tightly bound to the instantaneous abundances of the species, and the temperature evolution directly affects the dynamics of the system. This allows us to accurately describe the ISM out-of-equilibrium conditions, in particular when the gas is affected by shocks\footnote{With the term shocks we consider each hydrodynamic interaction in which $v_{\rm gas} \gg c_s$, and not only those that produce extremely hot gas with $T\gg10^5$~K.}; 
since such conditions are likely to occur for gas below $T=10^4$~K \citep[e.g.][]{bovino16}, this makes it hard to model them with a separate post-processing of the simulation results, especially since all these processes are tightly coupled.

Among the species included in our chemical network, we focus here on the line emission by two of the main coolants of the ISM, C$^+$ (whose cooling is dominated by the $157.7\rm\, \mu m$ transition) and O (for which two transitions are considered, at $63\rm\, \mu m$ and $146\,\rm \mu m$ respectively), similarly to what described in \citet{glover07} and \citet{grassi14}. To compute the intensity of the lines, \code{krome} assumes a statistical equilibrium of the atomic excited and ground states (including collisional excitation and de-excitation, spontaneous and stimulated emission, and photon absorption). At the low frequencies typical of the FIR lines, we can safely assume that the only relevant radiation is the Cosmic Microwave Background (CMB), that can alter the state populations and reduce the emission at low density, in particular in high-redshift galaxies above $z\sim 4.5$, when the CMB temperature reaches a few tens of K \citep{dacunha:2013,vallini15}. As an example, the equilibrium conditions can be written, for, e.g. a three-level ion like neutral O, as

\begin{equation}
\begin{bmatrix}
1 & 1 & 1 \\
T_{01} + T_{02} & - T_{10} & -T_{20} \\
-T_{01} & T_{10} + T_{12} & -T_{21} 

\end{bmatrix}
\begin{bmatrix}
n_0 \\ n_1 \\ n_2\\
\end{bmatrix}
=
\begin{bmatrix}
n_{\rm O} \\ 0 \\ 0\\
\end{bmatrix}
\end{equation}
where $T_{ij} = C_{ij} + B_{ij}I_{\nu_{ji}}$ and $T_{ji} = C_{ji} + A_{ji} + B_{ji}I_{\nu_{ji}}$, with $j>i = 0,1,2$ the state indices, $C_{ij}$ ($C_{ji}$) the collisional excitation (de-excitation) rate, $A_{ji}$ the spontaneous emission rate, $B_{ij}$ ($B_{ji})$ the absorption (stimulated emission) rate, and $I_{\nu_{ji}}$ the radiation intensity at the transition frequency $\nu_{ji}$. 
From the state populations obtained, we then compute the emissivity of each line as
\begin{equation}
    \Lambda_{\rm net} = \Lambda - \Gamma = n_j(A_{ji} + B_{ji}I_{\nu_{ji}}) - n_i B_{ij}I_{\nu_{ji}}
\end{equation}

\subsubsection{Post-processing the simulation with CLOUDY}\label{sub:sec:emission_cloudy}

An alternative to our non-equilibrium approach, that is typically employed for simulations where chemical species abundances are not directly evolved, is the post-processing with \code{cloudy}, that we describe in the following.
In \code{cloudy} models, {we assume that} each cell of the simulation is a homogeneous slab, which is then split in a multiple optically-thin layers for which species abundances, temperature, and line emission can be computed assuming photo-ionisation equilibrium conditions, and accounting for radiation absorption throughout the slab itself. Different choices can be made for the temperature throughout the slab, which can be kept fixed at a desired value or let free to evolve according to heating and cooling processes. Here we consider both cases, that we name CloudyFIX and CloudyVAR, respectively.

For both models, we generate a grid of \code{cloudy} models {using \citet{ferland17}} and interpolate the simulation data over the resulting multi-dimensional table, as a function of different parameters.
For the impinging flux, we assume the SED of {an} entire stellar population based on the updated \citet{bruzual03} models. {We assume the SED of a stellar population with $Z_{\star} = 0.5 \zsun$ and age $t_\star = 10\,Myr$, i.e. we select the stellar population that contributes most to the luminosity. The intensity of the radiation is then rescaled according to the local value of the Habing field found in the simulation.} Similarly to \citet{pallottini19}, we consider the full SED with its ionising component ($E_\gamma>13.6$~eV) only for particles having an ionising parameter $U>10^{-4}$, while a non-ionising radiation field is considered for the other particles.

In model CloudyVAR, {gas emission} is a function of the total gas density $n$, the gas metallicity $Z$, the UV radiation field $G$ (in units of the Milky-Way value $G_0$), and the total gas column density $N_{\rm gas}$.
{The grid of \code{cloudy} models covers a density range  $10^{-2}< n/{\rm cm}^{-3}< 10^{5}$, metallicities $10^{-3}< Z/\zsun< 10^{0.5}$, and intensities $10^{-1}< G/G_0< 10^{5}$. In the models we assume solar abundance ratios for the metals and linearly rescale the dust content {with metallicity assuming an 'ISM dust type'}. The \code{cloudy} calculation is stopped at $N_{\rm max}=10^{23}\rm cm^{-2}$, iterating to convergence}.

The gas temperature is allowed to vary throughout the slab according to the heating and cooling rates computed by \code{cloudy}, including photo-heating. This means that we lose consistency with the thermodynamic state determined in the simulation. Nevertheless, as we are limited by the resolution, and we cannot solve on-the-fly the photodissociation region (PDR) structure, this is the only choice able to exploit the chemistry and temperature changes within the PDR.

In model CloudyFIX, we force \code{cloudy} to maintain a constant temperature throughout the slab: effectively $T$ is considered as an additional parameter in the grid, that is interpolated using the values obtained from the simulation, {with value that ranges $1<T/{\rm K}<10^5$}. We notice that, in this case, the temperature of the slab is determined by full non-equilibrium evolution in the simulation, and does not necessarily correspond to the temperature one would get by employing \code{cloudy} tables for cooling and heating \citep[see][for details]{capelo18}.

This method has already been used in other studies \citep[e.g.][]{katz19}, as a way to mimic the presence of shocks \citep{egami18}, i.e. the photo-ionisation equilibrium assumption is only used to set the abundances (and the emission) at the chosen temperature \citep[see][for further discussion]{olsen18,pallottini19}.

{For CloudyVAR we use a 0.5 dex spacing for the variables ($n, Z, G$), yielding a total of $2\times 1344$ models where the factor two is due to having both ionizing and non-ionizing cases. For CloudyFIX we use a 1.0 dex spacing for the variables ($n, G$) and 0.5 for ($Z$, $T$), that sum up to a total of $2\times 4158$ models.} Note that, in our estimate of the emission line luminosity with \code{cloudy}, we do not assume any unresolved structure for the cells, unlike it is done in \citet{vallini18} and \citet{pallottini19}, to ensure a consistent comparison with the model `Krome' emission.

Although these approaches are all valid, each of them has its own advantages and weaknesses that should properly be acknowledged. In Table~\ref{tab:models}, we summarise all the differences among the models. 
Keep in mind that these differences are only related to how the instantaneous line emission is derived for every cell in the simulation at the desired redshift, and do not consider any time evolution, which is instead computed with \code{krome} according to the time-dependent non-equilibrium chemistry and cooling/heating processes.

In particular, although \code{cloudy} models exhibit a much higher accuracy in the treatment of the chemistry, ionisation equilibrium is assumed and does (CloudyFIX) or does not (CloudyVAR) take into account any dynamical effect onto the gas temperature. On the other hand, \code{krome} uses a simplified network and {is tied to the simulation resolution, i.e. it cannot model PDRs unless they are properly resolved in the simulation. However, it} naturally accounts for any deviations from ionisation equilibrium and dynamical effects naturally arising in the simulation.

\subsection{Predicted line emission}
\begin{figure*}
    CloudyFIX\hspace{4.2cm} CloudyVAR \hspace{4.2cm} Krome\\
    \centering
    \includegraphics[width=0.3\textwidth]{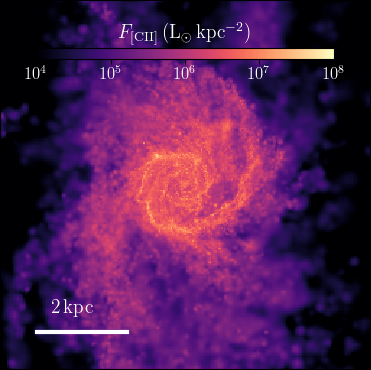}
    \includegraphics[width=0.3\textwidth]{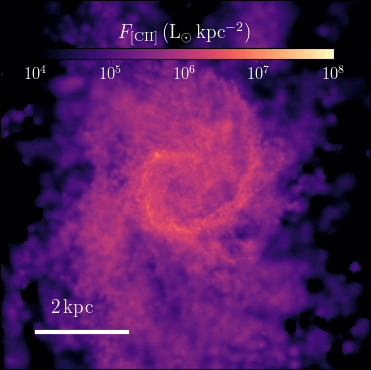}
    \includegraphics[width=0.3\textwidth]{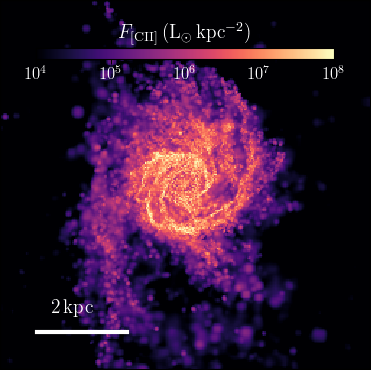}\\
    \includegraphics[width=0.3\textwidth]{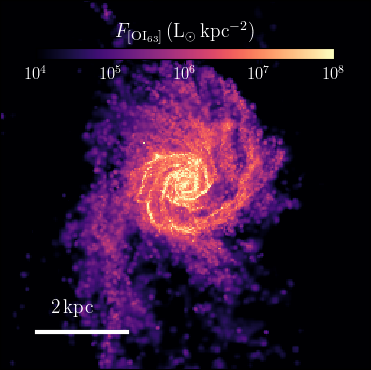}
    \includegraphics[width=0.3\textwidth]{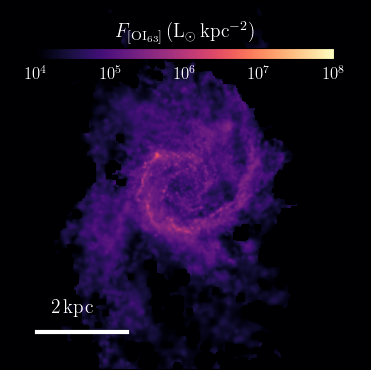}
    \includegraphics[width=0.3\textwidth]{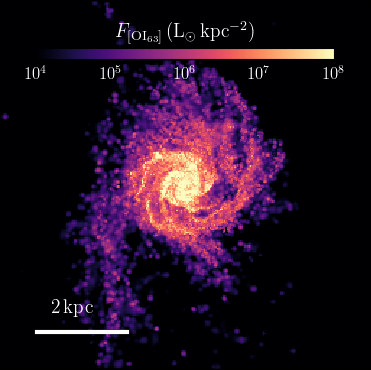}\\
    \includegraphics[width=0.3\textwidth]{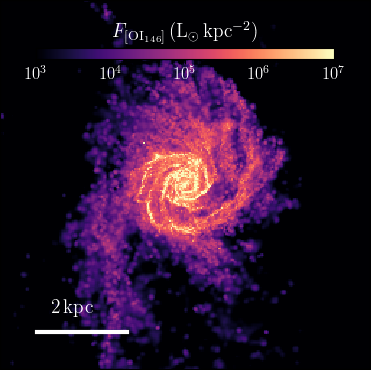}
    \includegraphics[width=0.3\textwidth]{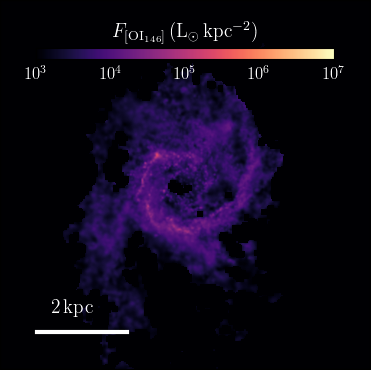}
    \includegraphics[width=0.3\textwidth]{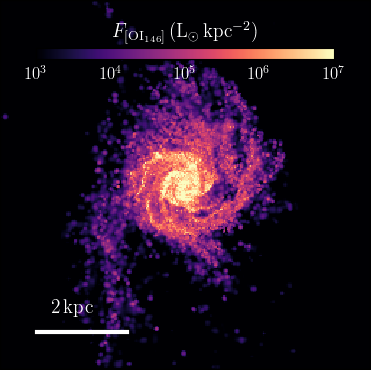}\\
    \includegraphics[width=0.3\textwidth]{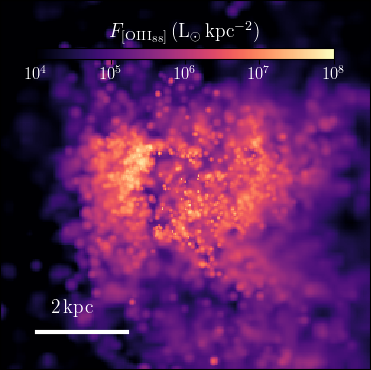}
    \includegraphics[width=0.3\textwidth]{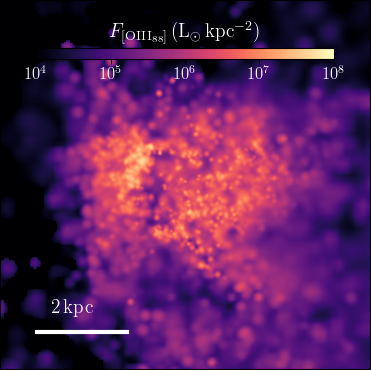}
	\rule{0.3\textwidth}{0cm}\\

    \caption{Emission maps of the galaxy at $z=6$ obtained from our different models, i.e. CloudyFIX (left panels), CloudyVAR (middle panels), and Krome (right panels). While for \cii~and \oiii~lines the assumption made for the cell temperature in \code{cloudy} is not important, resulting in moderate differences, for [OI] lines the emission varies by up to two orders of magnitude. For model Krome, we find still different results, but closer to model CloudyFIX than to CloudyVAR, because of the identical choice in the temperature of the cell.}
    \label{fig:emissionmap}
\end{figure*}

In Fig.~\ref{fig:emissionmap}, we compare the emission maps of \cii, \oia, and \oib~obtained with the three approaches reported in Table~\ref{tab:models}. We consider also \oiii, for which we only report the results obtained with \code{cloudy}, since O$^{++}$ was not included in our \code{krome} chemical network. The integrated luminosity over the maps are reported in Table~\ref{tab:emission}.

The \cii~maps show that \Cj{} emission from model `CloudyFIX' is stronger than that from `CloudyVAR', especially in the spiral arms and the galaxy nucleus, but no significant differences are observed in the outskirts of the galaxy. In model Krome, instead, the \cii~emission in the outskirts is slightly suppressed relative to the \code{cloudy} models, but is enhanced in the spiral arms of the galaxy. Overall, the total luminosity differs by a factor of $\sim 3$ between the smallest (CloudyVAR) and the largest value (Krome).

For \oia~and \oib~ the differences are much larger (a factor of $\sim 59$ and $\sim 76$, respectively). Models CloudyFIX and Krome predict a similarly strong emission from the dense gas in the spiral arms and a very weak emission from the low-density gas, always within a factor of two from each other (interestingly, \oib~is almost identical in the two models), whereas model CloudyVAR exhibits up to two orders of magnitude weaker emission across the entire galaxy. In the galaxy nucleus, this difference becomes larger, with [OI] emission being more centrally concentrated in models CloudyFIX and Krome, and almost negligible in CloudyVAR. Because of the stronger dependence of [OI] emission on the gas temperature, the differences we find can be easily explained with the typically lower temperatures predicted in the CloudyVAR model relative to those in the simulation.

For \oiii~emission, the differences between CloudyFIX and CloudyVAR are mild, with the emitting region being almost identical in the two cases. A possible explanation for this result is that, because of the high ionisation energy of \Oj~(35~eV), both SNe and ionising radiation can efficiently produce O$^{++}$; however, the gas shock-heated by SNe has typically low densities, while \oiii~traces denser gas. Hence, {the observed \oiii~emission can only be explained by the ionising radiation from young massive stars, which is typically absorbed in the outer layers of the slab, almost independently of the actual average gas temperature in the cells}.

\begin{table}
    \centering
    \caption{Total luminosity for the emission maps reported in Fig.~\ref{fig:emissionmap}.}
    \begin{tabular}{lcccc}
    \hline\hline
        Model & $L_{\rm [CII]_{158\mu}}$ & $L_{\rm [OI]_{63\mu}} $ & $L_{\rm [OI]_{146\mu}} $ &$L_{\rm [OIII]_{88\mu}} $\\
        & $\rm (L_\odot)$ & $\rm (L_\odot)$ & \rm $(L_\odot)$ & $\rm (L_\odot)$ \\ 
        \hline
        CloudyFIX & $10^{7.47}$ & $10^{7.77}$ & $10^{6.84}$ & $10^{7.66}$\\
        CloudyVAR & $10^{7.21}$ & $10^{6.24}$ & $10^{4.98}$ & $10^{7.65}$\\
        Krome & $10^{7.71}$ & $10^{8.01}$ & $10^{6.86}$ & $-$\\
    \hline\hline
    \end{tabular}
    \label{tab:emission}
\end{table}

\begin{figure*}
    \centering
    \includegraphics[width=0.95\textwidth]{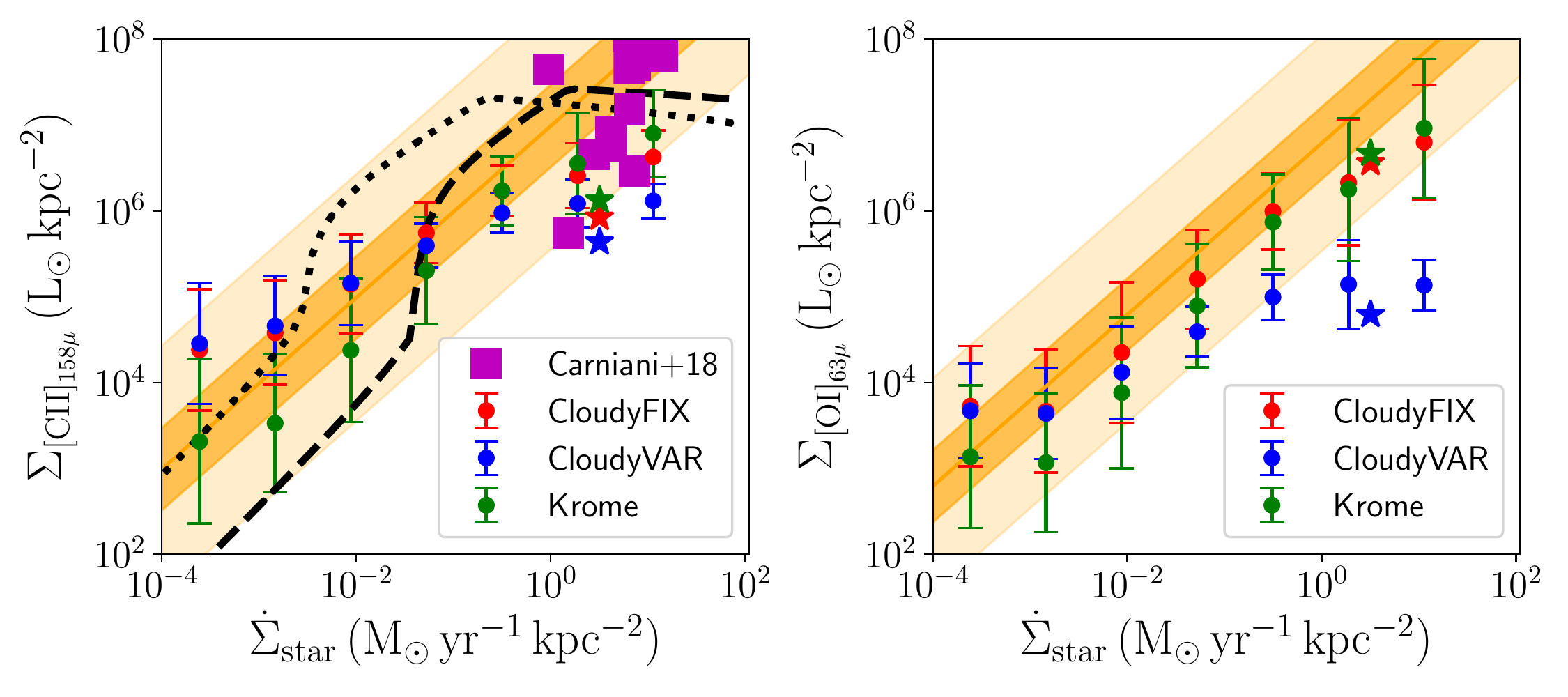}
    \caption{\cii--SFR (left-hand panel) and \oia--SFR (right-hand panel) correlations for the three different models applied to our simulation at $z=6$, compared with the results by \citet{delooze14}, shown as an orange solid line and the two shaded areas, the smallest corresponding to $1-\sigma$ and the largest to $3-\sigma$, and the theoretical model by \citet{ferrara:2019} for $Z=0.5\rm\, Z_\odot$ and $k_{\rm s}=1$ (black dotted line) and $k_{\rm s}=5$ (black dashed line). The coloured stars correspond to the average value for our models if the galaxy is observed at 2.5~kpc resolution. All models reproduce the \cii--SFR correlation reasonably well, with moderate differences in the slope. On the other hand, the \oia--SFR correlation can be reproduced only by Krome and CloudyFIX, with CloudyVAR predicting a strong deficit at high SFR surface densities. }
    \label{fig:correlations}
\end{figure*}

As a further comparison, we also investigate where the galaxy lies relative to the correlations between \cii~(and \oia) and the SFR observed at low (and high) redshift \citep{delooze14}. For this analysis, we extract the FIR fluxes from the maps in Fig.~\ref{fig:emissionmap}, degraded to 300~pc resolution, and the SFR surface density from an equivalent map of the intrinsic FUV emission of the galaxy, applying the conversion factor in \citet{salim07} for a Kroupa IMF. The results are shown in Fig.~\ref{fig:correlations}, where the left panel corresponds to the \cii--SFR correlation and the right panel to the \oia--SFR counterpart. The simulation data have been binned in 8 bins about 1-dex wide, and bins with less than 5 data points have been excluded to avoid any bias in the measure. The average value for the entire galaxy, is observed at 2.5~kpc resolution, is shown as coloured stars, using the colour corresponding to each model. The orange solid line and the two shaded areas correspond to the best-fit relation by \citet{delooze14} on the entire literature sample, assuming a perfectly linear slope, and its $1-\sigma$ (smallest) and $3-\sigma$ (largest) uncertainty. We also show the theoretical results by \citet{ferrara:2019} as black lines, obtained assuming a gas metallicity $Z=0.5\rm\, Z_\odot$ and a SFR lying $k_{\rm s}$ times off the Schmidt-Kennicutt relation, with $k_{s}=1$ (dotted line) and $k_{\rm s}=5$ (dashed line).

\begin{figure*}
    \centering
    \includegraphics[width=0.9\textwidth]{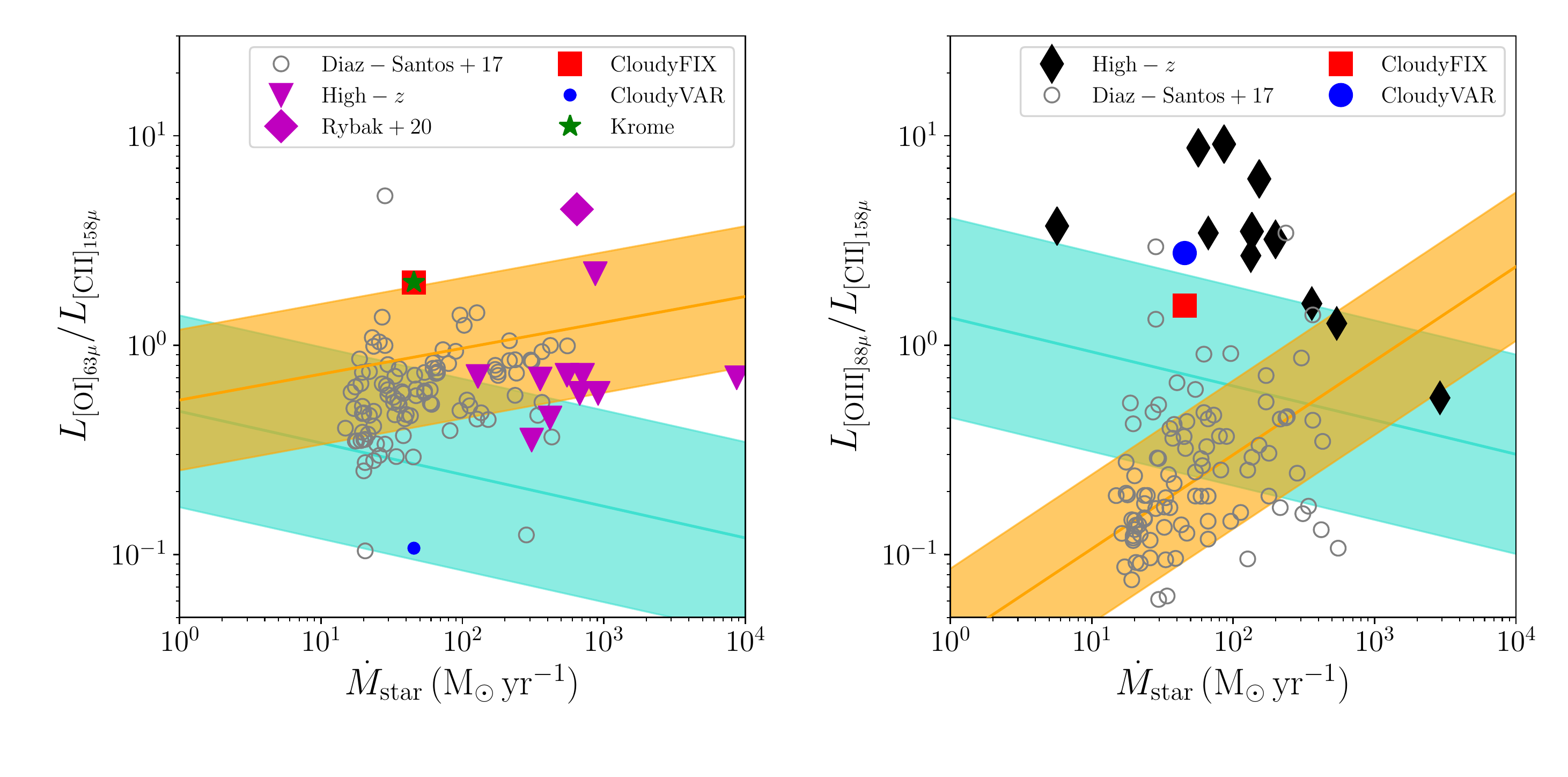}
    \caption{FIR line ratios for our different models, compared with local observations by \citet{delooze14} {of dwarf (cyan shaded area) and starburst galaxies} (orange shaded area), \citet{diaz-santos17} (grey circles), and high-redshift galaxies by \citet{carniani17,walter18,marrone18,hashimoto19,harikane19} (black diamonds), \citet{brisbin15,zhang18} (magenta triangles), and \citet{rybak20} (the magenta diamond in the left-hand panel). The red square corresponds to model CloudyFIX, the blue dot to CloudyVAR, and the green star to Krome. For the \oia--\cii~ratio, all models but CloudyVAR agree well with each other and observations. CloudyVAR, by predicting a very weak \oia~emission, results in a mild discrepancy with local data, and an even stronger one with other high-redshift data. In the right-hand panel, instead, CloudyVAR agrees better with observations, whereas CloudyFIX results in a factor of two lower ratio, smaller than typically observed values at the same SFR.}
    \label{fig:lineratio}
\end{figure*}

For the \cii~relation, our results (independent of the model considered) are slightly offset relative to the local relation, but consistent with observational results of high-redshift galaxies \citep[see, e.g.][]{carniani18}.  
Nevertheless, there are small differences in the slope of the relation. At low SFRs ($\dot{\Sigma}_{\rm star}\lesssim 1 \msun{\rm yr}^{-1}$), Krome well matches the observed slope  \citep[as already shown in][]{lupi20}, whereas the slope in models CloudyFIX and CloudyVAR is moderately shallower, with the data lying above the best-fit relation. At high SFRs ($\dot{\Sigma}_{\rm star}\gtrsim 1 \msun{\rm yr}^{-1}$), all models predict a saturation of the emission, consistent with the predictions by \citet{ferrara:2019}. Among the models, CloudyVAR exhibits the lowest saturation value, whereas CloudyFIX and Krome lie slightly above. This is due to the typically lower temperature in CloudyVAR for the higher-density cells, that likely correspond to the more star-forming regions of the galaxy. {Relative to the model by \citet{ferrara:2019}, Krome better follows the $k_{\rm s}=5$ case, although it does not show the enhanced \cii~luminosity around $\dot{\Sigma}_{\rm star}=0.1\rm\, M_\odot\, yr^{-1}\, kpc^{-2}$, and then saturates around $\dot{\Sigma}_{\rm star}=1\rm\, M_\odot\, yr^{-1}\, kpc^{-2}$. On the other hand, both \code{cloudy} models exhibit a higher luminosity at low SFRs, more consistent with the $k_{\rm s}=1$ case, and then deviate towards the $k_{\rm s}=5$ case, finally saturating at a SFR similar to that observed for Krome. The average values, that are dominated by the central patches where the relation saturates, exhibit a deficit relative to the local relation, but are still reasonably consistent with \citet{carniani18}.}

{For [OI] lines, the differences are more noticeable. Models CloudyFIX and Krome are quite similar, and follow reasonably well the observed slope (also when the average values are considered).
Model CloudyVAR produces instead a much lower emission from dense gas, resulting in a very shallow relation and significant deviations at large SFR surface densities, also reflected in the average value.}

Observationally, lines ratios such as \oia/\cii~ and \oiii/\cii~ have been widely use to investigate the ISM properties. {In particular, \cii~ is emitted in both cold neutral medium and PDRs, \oia~ by dense gas and warm PDRs (as long as other mechanisms like mechanical or X-ray heating do not dominate), and \oiii~by the ionised gas near stellar sources. Hence, the \oia--\cii~ ratio represents a good tracer of the typical ISM density and PDR temperature, whereas the \oiii--\cii~ gives us information about the ionisation parameter in the ISM and/or the PDR filling factor.}  Fig.~\ref{fig:lineratio} shows these ratios for our galaxy, compared to high-redshift observations. 
{In the left-hand panel, we see that all models but CloudyVAR appear moderately consistent with local starbursts by \citet{delooze14} and \citet{diaz-santos17}, with the ratio being in the upper end of the observed range. Relative to high-redshift galaxies, our simulation lies in between the $z\sim 1-4$ and the $z\sim 6$ data . CloudyVAR, on the other hand, because of the extremely weak \oia~ emission, results in an extremely low \oia/\cii~ ratio, lying at the lower end of the local dwarf region. The \oiii/\cii~ ratio varies by about a factor of two between the two \code{cloudy} models, with CloudyVAR in this case being closer to the observed high-redshift data with respect to CloudyFIX. Compared to the local relations by \citet{delooze14}, {our simulation (for both models) agrees better with the dwarf galaxy correlation rather than with starbursts by \citet{delooze14}}, although it is still compatible with the highest data points of the LIRG sample by \citet{diaz-santos17}.}

\subsection{Phase diagrams and ion abundances}

\begin{figure*}
    \centering
    \includegraphics[width=\textwidth,trim=0cm 2cm 0 2cm,clip]{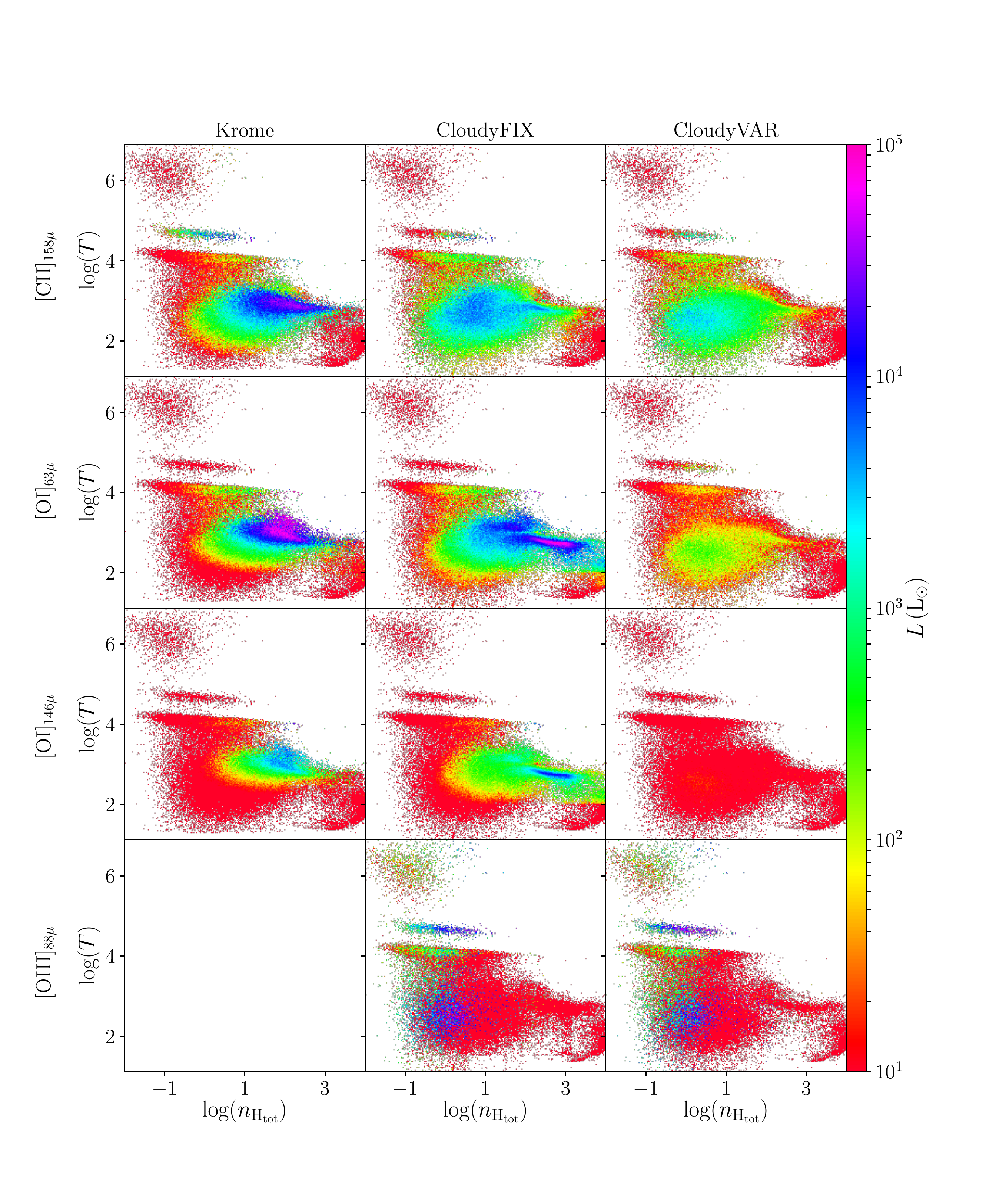}
    \caption{
    Density--temperature diagram for our three models, weighted by the FIR line luminosity. The first row corresponds to \cii, the {second} one to \oia, {the third to \oib,} and the bottom one to {\oiii}, with Krome shown in the left-most column, CloudyFIX in the middle one, and CloudyVAR in the right-most one. \cii~ luminosity is reasonably similar in all models, with the only noticeable differences being a higher peak in Krome at high density, and a globally lower emission in CloudyVAR. For [OI] lines, CloudyVAR results in extremely low emission, consistent with the very low line ratio in Fig.~\ref{fig:lineratio}, whereas Krome and CloudyFIX give more similar results, with the stronger difference resulting from the high-density region. At low densities, \code{cloudy} models give somewhat larger luminosities for all lines compared to Krome. A similar discrepancy is found at very high densities, where no emission is obtained with Krome, whereas CloudyFIX produces a non-negligible fraction of the line luminosity. {For \oiii, no noticeable differences can be appreciated between the two \code{cloudy} models, where most of the emission comes from ionised gas above $T\sim 10^4$~K.}}
    \label{fig:emissionlines}
\end{figure*}

In order to better constrain the origin of the differences in luminosity, we inspect the typical thermodynamic conditions of the gas responsible for the emission. In Fig.~\ref{fig:emissionlines}, we show the density--temperature diagrams for our galaxy at $z=6$, weighted by the \cii~ (top row), \oia~ (second row),{ \oib~ (third row), and \oiii~(bottom row)} luminosity.

For \cii, all models are reasonably similar, apart from the warm high-density region where Krome predicts a factor of a few higher luminosity compared to CloudyFIX and CloudyVAR. Because of the variable temperature across the slab, CloudyVAR predicts the lowest luminosity among the models, but still reasonably consistent with them (within a factor of $\sim3$). Interestingly, \code{cloudy} models give a moderately higher luminosity at low densities relative to Krome, likely because of the assumption of ionisation equilibrium, which is harder to reach \citep[see][for a discussion]{oppenheimer13,richings14,bovino16}. 

For [OI] lines, instead, the differences are much larger, with Krome and CloudyFIX exhibiting an emission peak in warm, high-density gas compared to CloudyVAR. As for \cii, Krome shows the highest luminosity in this regime (a factor of a 2-3 higher compared to CloudyFIX). However, the [OI] emission in CloudyFIX at very high densities ($n_{\rm H_{tot}}>100\rm\, cm^{-3}$) is somewhat larger than that in Krome, suggesting that [OI] emission in this regime is stronger in photoionisation equilibrium conditions.
CloudyVAR, instead, exhibits extremely low luminosities for [OI] lines, likely because the temperature within the slab drops quickly to low values where O is not efficiently excited.

In order to understand the reason for the higher luminosity in Krome for gas in the range $10\rm\, cm^{-3}<n _{\rm H_{tot}}<1000\rm\, cm^{-3}$ compared to CloudyFIX, we checked the {abundance of CO in the corresponding density range with \code{cloudy}}, finding that a non-negligible part of C and O are indeed bound into CO \citep[see, e.g,][for a discussion]{glover07}, not included in our chemical network.
Although, in principle, such a difference could also arise because of the \oia~ line becoming {optically} thick in the inner layers of the cell/slab (where we reach $A_{\rm V}\sim 3-4$; see, \citealt{kaufman1999} for a discussion), 
an effect not modelled in Krome, this discrepancy should increase at even higher densities, where $A_{\rm V}$ is larger \citep[see, e.g.,][]{grassi17}. However, this is not seen in our case, where at $n_{\rm H_{tot}}\gtrsim 10^3\rm\, cm^{-3}$ CloudyFIX predicts larger luminosities compared to Krome, we can safely assume this effect does not play a significantly role on our results. At low-densities, instead, the higher \cii~ and \oia~ luminosities in Krome are due to the higher ionisation states, not included in our network (see, for instance, the hot gas where C$^{++}$ should form).

{Finally, for \oiii, we see that the emission is almost identical between CloudyVAR and CloudyFIX, with most of it coming from ionised gas above $T\sim 10^4$~K. Since in CloudyVAR the temperature evolves according to the depth within the slab, this result suggests that most of the emission is produced in the first layers of the slab, where ionising photons have not been totally absorbed yet, and the temperature distribution in the deeper layers does not significantly affect it.}

\begin{figure*}
    \centering
    \includegraphics[width=0.9\textwidth,trim=0cm 2cm 0 2cm,clip]{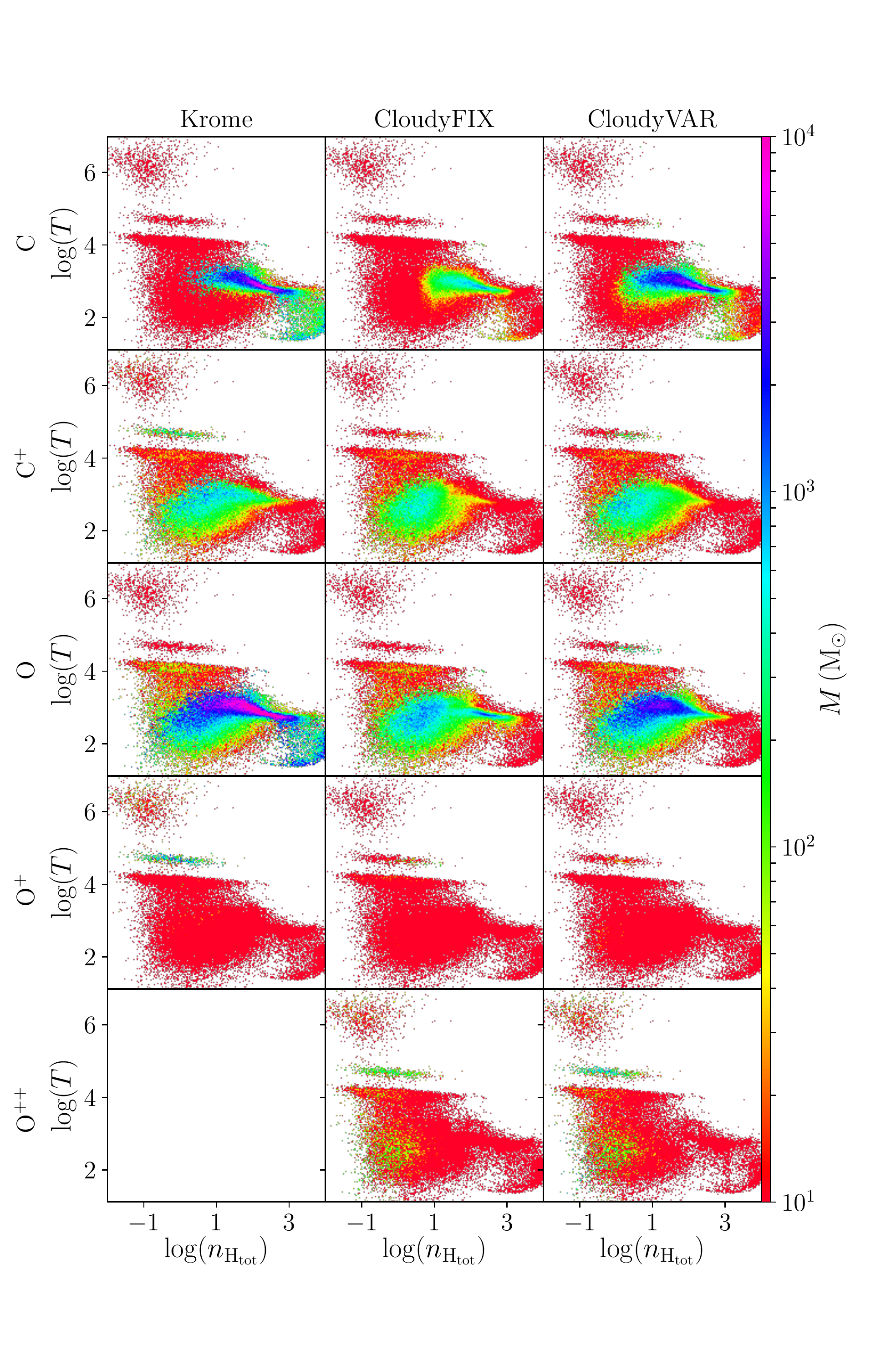}
    \caption{Same as Fig.~\ref{fig:emissionlines}, weighted by the ion abundance. All models look similar in terms of \Oj~and \Cj~masses, with mild differences due to the missing CO in Krome and to possible small deviations from photoionisation equilibrium. Neutral species are instead more affected by the different thermodynamic assumptions in the models. Krome, because of the missing CO, produces too much O and C at high densities ($n_{\rm H,tot}>10\rm\, cm^{-3}$), although similar abundances for C are found in CloudyVAR as well. CloudyFIX, instead, results in lower abundances of these ions, because of the more efficient conversion to CO at $T\sim 500-1000$~K. {For O$^{++}$, CloudyVAR and CloudyFIX result in about the same mass, with most of O$^{++}$ found in gas at $T>10^4$~K.}}
    \label{fig:masscompare}
\end{figure*}

To assess whether these differences arise from the different ion abundances or are linked to the determination of the gas temperature and the photoionisation conditions, we show in Fig.~\ref{fig:masscompare} phase diagrams weighted by the C, \Cj, O, and \Oj~abundances. The first column shows the mass of each species in our simulation, as obtained by \code{krome}, whereas the second and third ones correspond to models CloudyFIX and CloudyVAR, respectively.

C and \Cj~behave in a reasonably similar way in all models, with neutral C dominating the high-density tail of the gas, even at $T\sim1000$~K, and \Cj~being the most important ion for lower density gas, at all temperatures up to a few $10^4$~K. While the agreement among the models for \Cj~is very good, with only mild differences consistent with those observed in the luminosity diagrams (within a factor of two), C is more significantly affected by the model choice, with total masses that differ by up to a factor of 7 (between the two \code{cloudy} models).
In this case, we observe an opposite behaviour compared to Fig.~\ref{fig:emissionlines}, with Krome and CloudyVAR showing similar abundances, whereas CloudyFIX predicts a lower abundance of neutral C{. This difference can be easily explained by considering the formation channels of H$_2$ and CO in the ISM. At $T\sim 100$~K (typical of the inner regions of the slab in CloudyVAR), most of the hydrogen is in H$_2$, and this suppresses the formation of CH, which is then turned into CO. On the other hand, at temperatures of $\sim 1000$~K (which are more frequent in CloudyFIX), atomic H is more abundant, and CH (and then CO) can form more efficiently.} 

\Oj~is abundant only in warm/hot gas above $10^4$~K in Krome, and is almost completely missing in CloudyFIX and CloudyVAR, due to its conversion to O$^{++}$ or even higher ionisation states\footnote{Since both in \code{krome} and \code{cloudy} we assume solar ratios for the metal abundances, the total amount of oxygen nuclei is expected to be the same in all the three models, and the only differences are due to the relative abundance of different ionisation states and/or the formation of oxygen-bearing molecules.}. Neutral O is instead predominant everywhere in the galaxy ISM in all models. For O, the agreement is reasonably good among all models, with the largest differences appearing in the CO-forming region {($10\lesssim n_{\rm H_{tot}}/({\rm cm^{-3}})\lesssim 10^3$ and $T\sim 3000$~K)}. Surprisingly, also in this case Krome agrees better with CloudyVAR than with CloudyFIX. 

We conclude that, rather than the actual metal ion abundances, temperature plays a major role in determining the emission. This is because $T$ regulates both {the species abundances and the excitation cross sections, hence affecting the line emission efficiency. As a consequence, even lower abundances of the emitting ion can result in a larger luminosity.}

\section{Caveats}
\label{sec:caveats}
Before moving to the conclusions, there are a few important caveats that should be considered.

\subsection{Uncertainties of the KROME model}  

First, our non-equilibrium approach, by not including the formation and dissociation of CO, might overestimate the emission of neutral atoms like O, especially at high density.
Although the inclusion of CO could improve significantly the prediction capabilities of our approach, the additional cost of a CO network would have made our cosmological simulations much more computationally expensive to perform (we would need at least $\sim 300$ reactions as in \citealt{grassi17} against the $\sim 85$ employed here), and its inclusion is deferred to future studies.
{Another possible issue is related to the \oia~ emission, because of the pathological nature of this line, which can often become optically thick \citep{kaufman1999}, a regime not included in our cell-average approach with \code{krome}. Nevertheless, the optical depth in our \oia-luminous cells is not extremely large, and this effect is most likely marginal relative to CO conversion.}

Secondly, the chemical network adopted in this study does not include highly ionised species like C$^{++}$ and O$^{++}$. These species could in principle decrease the amount of \Cj and \Oj in the ISM, and slightly suppress the emission, giving results less accurate than with \code{cloudy}. However, our analysis showed that the typical conditions for double (at least) ionisation, i.e. very hot gas or the presence of hard radiation, are not typical of the ISM of star-forming galaxies{, with the only exception of HII regions, where ionising radiation can efficiently power \oiii~ emission. Hence, given that the amount of ionised mass in HII regions is subdominant compared to the entire ISM, and in these regions our network would simply predict more O$^+$ (as can be actually seen in Fig.~\ref{fig:emissionlines}), we expect this approximation not to significantly affect our conclusions about [OI] and \cii~line emission.}

Third, chemistry is limited by the simulation resolution when \code{krome} is employed, and any possible sub-grid distribution cannot be accounted for.

\subsection{Uncertainties of the CLOUDY models  }  

Caveats are also related to the post-processing performed with \code{cloudy}. The first and most important is that the radiative flux extracted from the cell and passed to \code{cloudy} is already processed in the simulation, hence could be already attenuated compared to the intrinsic flux impinging on the cell and actually affecting the chemistry.

Secondly, if we let \code{cloudy} recompute the temperature within the slab, we risk neglecting gas shocks and getting a temperature that is not fully consistent with the one in the simulation. Although the temperature distribution through the slab could be more realistic in equilibrium conditions, this is not consistent with the hydrodynamics we rely upon.
If, instead, we keep the temperature constant through the slab at the value obtained in the simulation, we are more consistent with the simulation, but we do not properly consider how the gas temperature would evolve through the slab because of radiation absorption.  
Thirdly, we have to keep in mind that \code{cloudy} is reliable as long as the gas is in photo-ionisation equilibrium and no significant deviations from occur in the gas \citep{richings14,richings16}, and this strongly depends on the dynamical evolution of the system under scrutiny, so particular care should be taken.
At last, we have to consider that in most simulations, where chemistry is not coupled with the hydrodynamics, cooling and heating are computed according to \code{cloudy} calculations including (or not) a uniform UV background, but neglecting the interstellar radiation field, which is instead included in subsequent emission line calculations, producing possibly inconsistent results.

{\subsection{Additional uncertainties}
Obviously, the comparison presented in this work only covers a small part of the explorable parameter space. Among the additional variations that can be explored there are \textit{(i)} stellar-related properties like stellar spectra, stellar yields, SN rates/energetics, and the inclusion of first population stars and pair-instability SNe;
\textit{(ii)} changes in the sub-grid modelling (e.g. star formation, chemical abundance ratio, UV background, enhanced/suppressed feedback effects);
\textit{(iii)} the inclusion of a sub-grid structure for the cell, as done in \citet{pallottini19}, although we stress that this approach can only be employed with post-processing, since a full description of the non-equilibrium chemistry on unresolved scales during the simulation is not feasible; \textit{(iv)} changes in the \code{cloudy} parameters.
All of these effects could significantly change the galaxy evolution, hence the emission. However, such a complete exploration is still prohibitive at the moment, especially because of the high computational cost of the non-equilibrium chemistry simulations.
}
\section{Discussion and conclusions}\label{sec:conclusions}
In this work, we presented a state-of-the-art cosmological simulation of a star forming galaxy at high-redshift that includes, for the first time, on-the-fly radiative transfer (in ten bins ranging from 0.75~eV up to 1 keV) and non-equilibrium chemistry for the primordial (H, H$^+$, H$^-$, He, He$^+$, He$^{++}$, H$_2$, and H$_2^+$) and the main metal species (C, C$^+$, O, O$^+$, Si, Si$^+$, and Si${++}$).  
Our subgrid model is an improved version of the one presented in \citet{lupi20}. 
Our simulation predicts an evolved galaxy with already roughly solar metallicity and well developed stellar and gaseous discs, similar to previous results \citep{pallottini19,katz19}.
Thanks to the extremely high spatial ($\sim 3$~pc) and mass ($2\times 10^4\rm\, M_\odot$ per baryonic particle) resolution, and the coupling with \code{krome} \citep{grassi14}, we were able to investigate the impact of non-equilibrium thermochemistry on the system evolution and to assess its effect on the Far Infrared (FIR) emission lines.
We compared our non-equilibrium treatment with \code{cloudy} models applied in post-processing on the simulation \citep{vallini:2017,pallottini19}, making different assumptions about the thermal structure within each cell. This analysis allowed us to constrain the uncertainty in the predicted FIR line luminosity and flux spatial distribution {coming from different theoretical models}.

Our results show that, despite the huge conceptual differences in the three approaches we considered, the \cii~ luminosity is not strongly affected by our choice (up to a factor of $3$); hence, the estimates can be considered reliable and be used as a tool to infer the ISM properties in observed galaxies. As a consequence, we can safely consider \cii~ as a good tool to constrain the physics included in the simulations, that could help us better constraining our understanding of galaxy formation.

On the other hand, [OI] lines are much more affected by line model choice, which translates into a less predictive power of the hydrodynamic simulations. {For instance, the \oia/\cii~ ratio inf Fig.~\ref{fig:lineratio} either predicted, according to the chosen model, a dense/warm ISM or a low-density and cold one.} Interestingly, this makes [OI] lines crucial to disentangle the models and assess which assumptions about the thermodynamic state of the gas better reproduce the ISM of high-redshift galaxies.

In addition to \cii~ and [OI] lines, \oiii~ represents an important tracer of the ionisation state of the gas around stellar sources, and is becoming more and more important in observations. Although not included in our current chemical network, we compared the two \code{cloudy} approaches, finding that \oiii~ emission is not significantly affected by the temperature assumption, but only by the ionising flux reaching the cloud. {The reason is that, in the regions where \oiii~is efficiently emitted, the gas temperature is mainly set by the ionising radiation, hence it is consistent between the simulation, where radiation transport is directly coupled to hydrodynamics, and \code{cloudy}, where it is only applied in post-processing. This makes the \oiii~ line a powerful tool to investigate the ionisation state of the ISM, and will be included in a new chemical network we are currently working on, that we plan to employ in future works.} However, particular care needs to be taken with the \oiii/\cii~ ratio, since high values could either support the case of a higher ionisation parameter/lower covering fraction of PDRs, or simply a less efficient \cii~ emission.

Concluding, the good agreement between Krome and CloudyFIX we found seems to suggest that gas in our simulated galaxy is not always far from equilibrium, hence that tabulated \code{cloudy} calculations represent an accurate and `cheaper' alternative to on-the-fly chemistry. Nevertheless, we must notice that the input temperature in \code{cloudy} has been computed via a self-consistent non-equilibrium modelling of the chemistry of both primordial and metal species and all the relevant heating/cooling processes of the gas, and not via equilibrium tables where the effect of a variable radiation flux is not accounted for. This means that the temperature distribution we consider is more accurate than that we would have obtained employing equilibrium tables, and, as a consequence, the emission line calculations as well. Therefore, we stress that a proper inclusion of non-equilibrium chemistry calculations in simulations as done by \code{krome} is crucial to get a better understanding of the interplay between microphysics and dynamics, and of galaxy formation and evolution in general, as also shown by \citet{richings16} and \citet{capelo18}.

\section*{Acknowledgements}
We thank the anonymous referee for useful suggestions that helped to improve the quality of the manuscript.
AL, AF, and SC acknowledge support from the European Research Council No. 740120 `INTERSTELLAR'. This work reflects only the authors' view and  the  European Research Commission is not responsible for information it contains.
Support from the Carl Friedrich von Siemens-Forschungspreis der Alexander von Humboldt-Stiftung Research Award is kindly acknowledged (AF).
SB thanks for funding through PCI Redes Internacionales project number REDI170093, Conicyt PIA ACT172033 and BASAL Centro de Astrof\'isica y Tecnolog\'ias Afines (CATA) PFB-06/2007

\section*{Data Availability Statements}
The data underlying this article will be shared on reasonable request to the corresponding author.


\bibliographystyle{mnras}
\bibliography{Biblio} 




\appendix
\section{The CMB attenuation}

\begin{figure}
    \centering
    \includegraphics[width=\columnwidth]{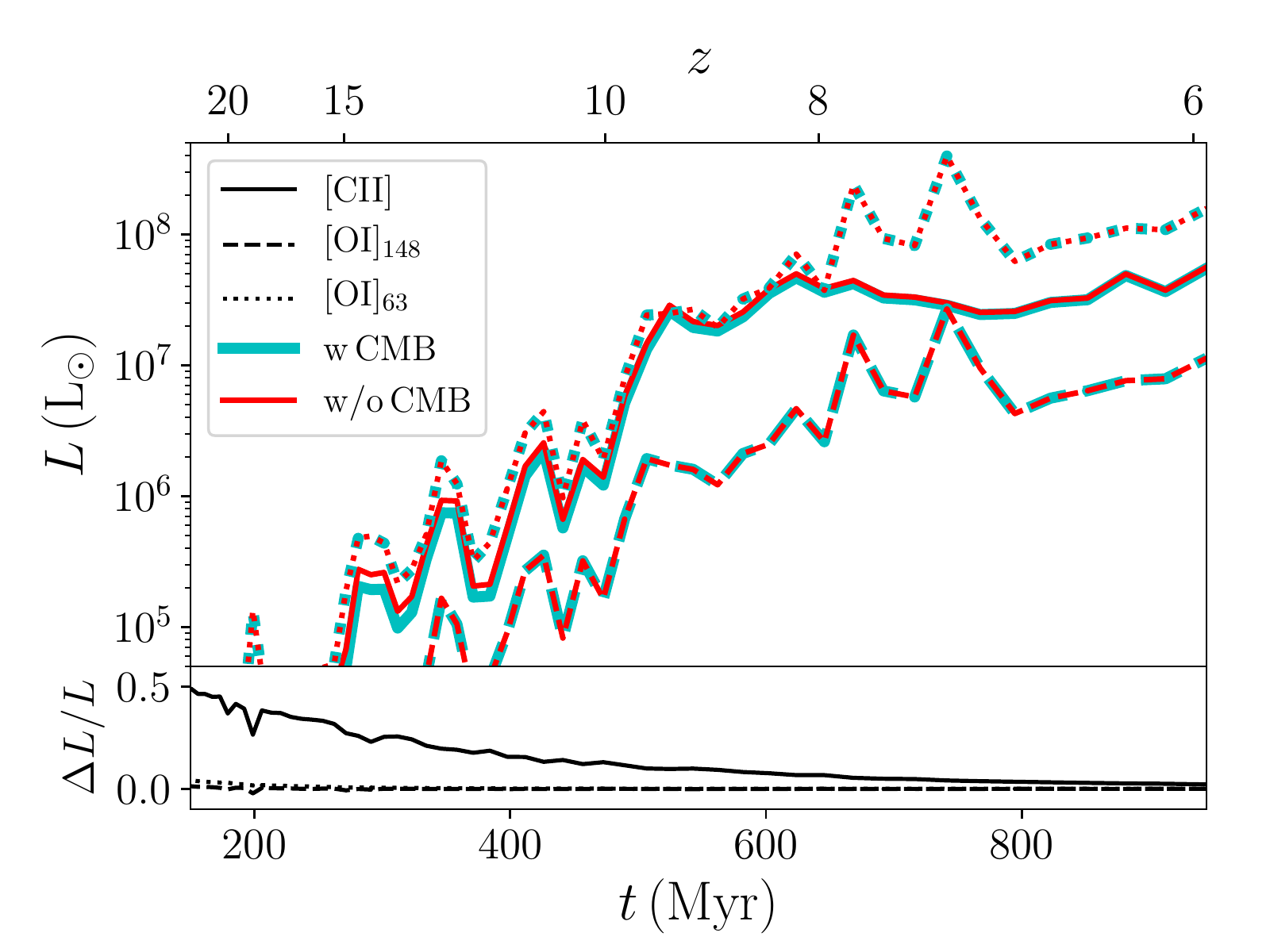}
    \caption{Cumulative line emission as a function of redshift for our galaxy, with and without including the CMB effect. In the top panel, thick/thin lines correspond to the cases with/without the CMB, and solid, dashed and dotted styles correspond to \cii,\oib, and \oia, respectively. In the bottom panel, we show the relative difference due to the CMB. We can clearly see that the effect becomes large for \cii~ only at very high redshift, and it does not exceed 10 per cent between $z=6$ and $z=9$. [OI] lines, instead, are not affected at all by the CMB in the typical conditions we find in our simulation, but the results could significantly change with the gas temperature, since [OI] emission is more strongly influenced by temperature variations than \cii.}
    \label{fig:cmbatt}
\end{figure}

At high redshift, CMB represents a strong diffuse background at the same frequency of the FIR lines, over which the signal is detected as contrast \citep{dacunha:2013,vallini15}. As detailed in the main text, in our analysis we assume that the state populations can be affected by CMB photons. However, recent results have suggested that this effect could be much smaller than expected, except for very-low density gas where emission would be low anyway \citep{pallottini17a,arata20}. Here, we test this idea by comparing the line flux obtained by \code{krome} with and without including the CMB effect. In Fig.~\ref{fig:cmbatt}, we show the evolution of the FIR lines emission as a function of redshift, with (thick lines) and without (thin lines) the CMB effect. In the top panel, we show the total luminosity for \cii~ (solid lines), \oia~ (dashed lines), and \oib~ (dotted lines), whereas in the bottom one we show the relative difference for each line. \cii~ is mildly affected, and the CMB impact increases with increasing redshift, up to 50 per cent at extremely high redshift. For [OI] lines, instead, CMB has negligible effect at all redshifts.

\bsp	
\label{lastpage}
\end{document}